

\input amssym.tex 

\def\unredoffs{}
\tolerance=1000\hfuzz=2pt
\catcode`\@=11 
\ifx\hyperdef\UNd@FiNeD\def\hyperdef#1#2#3#4{#4}\def\hyperref#1#2#3#4{#4}\def\href#1#2{#2}\fi
\magnification=1200\unredoffs\baselineskip=16pt plus 2pt minus 1pt
\def\Date#1{\vfill\leftline{#1}\tenpoint\supereject%
\footline={\hss\tenrm\hyperdef\hypernoname{page}\folio\folio\hss}}%

{\count255=\time\divide\count255 by 60 \xdef\hourmin{\number\count255}
 \multiply\count255 by-60\advance\count255 by\time
 \xdef\hourmin{\hourmin:\ifnum\count255<10 0\fi\the\count255}
}
\def\date{\number\day.\number\month.\number\year\ at \hourmin}


\def\nolabels{\def\wrlabeL##1{}\def\eqlabeL##1{}\def\reflabeL##1{}}
\def\writelabels{\def\wrlabeL##1{\leavevmode\vadjust{\rlap{\smash%
{\line{{\escapechar=` \hfill\rlap{\sevenrm\hskip.03in\string##1}}}}}}}%
\def\eqlabeL##1{{\escapechar-1\rlap{\sevenrm\hskip.05in\string##1}}}%
\def\reflabeL##1{\noexpand\llap{\noexpand\sevenrm\string\string\string##1}}}
\nolabels

\global\newcount\secno \global\secno=0
\global\newcount\meqno \global\meqno=1
\def\s@csym{}

\def\newsec#1\par{\global\advance\secno by1%
{\toks0{#1}\message{(\the\secno. \the\toks0)}}%
\global\subsecno=0\eqnres@t\let\s@csym\secsym\xdef\secn@m{\the\secno}\noindent
{\bf\hyperdef\hypernoname{section}{\the\secno}{\the\secno.} #1}%
\writetoca{{\string\hyperref{}{section}{\the\secno}{\bf \the\secno\quad}} {\bf #1}}\par%
\nobreak\medskip\nobreak\noindent\ignorespaces}
\def\eqnres@t{\xdef\secsym{\the\secno.}\global\meqno=1\bigbreak\bigskip}
\def\sequentialequations{\def\eqnres@t{\bigbreak}}\xdef\secsym{}

\global\newcount\subsecno \global\subsecno=0
\def\subsec#1\par{\global\advance\subsecno by1%
{\toks0{#1}\message{(\s@csym\the\subsecno. \the\toks0)}}%
\global\subsubsecno=0%
\ifnum\lastpenalty>9000\else\bigbreak\fi
\noindent{\it\hyperdef\hypernoname{subsection}{\secn@m.\the\subsecno}%
{\secn@m.\the\subsecno.} #1}\writetoca{\string\hskip1.45cm
{\string\hyperref{}{subsection}{\secn@m.\the\subsecno}{\secn@m.\the\subsecno.}}
{#1}}\par\nobreak\medskip\nobreak\noindent\ignorespaces}

\global\newcount\subsubsecno \global\subsubsecno=0
\def\subsubsec#1\par{\global\advance\subsubsecno by1%
{\toks0{#1}\message{(\secn@m.\the\subsecno.\the\subsubsecno. \the\toks0)}}%
\global\subsubsubsecno=0%
\ifnum\lastpenalty>9000\else\bigbreak\fi
\noindent{\it\hyperdef\hypernoname{subsubsection}{\secn@m.\the\subsecno\the\subsubsecno}%
{\secn@m.\the\subsecno.\the\subsubsecno.} #1}
\par\nobreak\medskip\nobreak\noindent\ignorespaces}

\global\newcount\subsubsubsecno \global\subsubsubsecno=0
\def\subsubsubsec#1\par{\global\advance\subsubsubsecno by1%
{\toks0{#1}\message{(\secn@m.\the\subsecno.\the\subsubsecno.\the\subsubsubsecno \the\toks0)}}%
\ifnum\lastpenalty>9000\else\bigbreak\fi
\noindent{\it\hyperdef\hypernoname{subsubsection}{\secn@m.\the\subsecno\the\subsubsecno\the\subsubsubsecno}%
{\secn@m.\the\subsecno.\the\subsubsecno.\the\subsubsubsecno.} #1}%
\par\nobreak\medskip\nobreak\noindent\ignorespaces}


\def\newnewsec#1#2\par{\global\advance\secno by1%
{\toks0{#2}\message{(\secn@m. \the\toks0)}}%
\global\subsecno=0\global\subsubsecno=0\eqnres@t\let\s@csym\secsym\xdef\secn@m{\the\secno}\noindent
\ifnum\lastpenalty>9000\else\bigbreak\fi
\noindent{\bf\hyperdef\hypernoname{section}{\secn@m}{\secn@m.} #2}%
\writetoca{{\string\hyperref{}{section}{\the\secno}{\bf \the\secno\quad}} {\bf #2}}
\DefWarn#1%
\xdef#1{\noexpand\hyperref{}{section}{\the\secno}%
{\the\secno}}\writedef{#1\leftbracket#1}\wrlabeL{#1=#1}%
\par\nobreak\medskip\nobreak\noindent\ignorespaces}

\def\newsubsec#1#2\par{\global\advance\subsecno by1%
{\toks0{#2}\message{(\secn@m.\the\subsecno. \the\toks0)}}%
\global\subsubsecno=0%
\ifnum\lastpenalty>9000\else\bigbreak\fi
\noindent{\it\hyperdef\hypernoname{subsection}{\secn@m.\the\subsecno}%
{\secn@m.\the\subsecno.} #2}
\DefWarn#1%
\xdef#1{\noexpand\hyperref{}{subsection}{\secn@m.\the\subsecno}%
{\secn@m.\the\subsecno}}\writedef{#1\leftbracket#1}\wrlabeL{#1=#1}%
\writetoca{\string\hskip1.45cm
{\string\hyperref{}{subsection}{\secn@m.\the\subsecno}{\secn@m.\the\subsecno.}}
{#2}}%
\par\nobreak\medskip\nobreak\noindent\ignorespaces}

\def\newsubsubsec#1#2\par{\global\advance\subsubsecno by1%
{\toks0{#2}\message{(\secn@m.\the\subsecno.\the\subsubsecno. \the\toks0)}}%
\global\subsubsubsecno=0%
\ifnum\lastpenalty>9000\else\bigbreak\fi
\noindent{\it\hyperdef\hypernoname{subsubsection}{\secn@m.\the\subsecno\the\subsubsecno}%
{\secn@m.\the\subsecno.\the\subsubsecno.} #2}
\DefWarn#1%
\xdef#1{\noexpand\hyperref{}{subsubsection}{\secn@m.\the\subsecno.\the\subsubsecno}%
{\secn@m.\the\subsecno.\the\subsubsecno}}\writedef{#1\leftbracket#1}\wrlabeL{#1=#1}%
\par\nobreak\medskip\nobreak\noindent\ignorespaces}

\def\newsubsubsubsec#1#2\par{\global\advance\subsubsubsecno by1%
{\toks0{#2}\message{(\secn@m.\the\subsecno.\the\subsubsecno.\the\subsubsubsecno \the\toks0)}}%
\ifnum\lastpenalty>9000\else\bigbreak\fi
\noindent{\it\hyperdef\hypernoname{subsubsection}{\secn@m.\the\subsecno\the\subsubsecno\the\subsubsubsecno}%
{\secn@m.\the\subsecno.\the\subsubsecno.\the\subsubsubsecno.} #2}
\DefWarn#1%
\xdef#1{\noexpand\hyperref{}{subsubsubsection}{\secn@m.\the\subsecno.\the\subsubsecno.\the\subsubsubsecno}%
{\secn@m.\the\subsecno.\the\subsubsecno.\the\subsubsubsecno}}\writedef{#1\leftbracket#1}\wrlabeL{#1=#1}%
\par\nobreak\medskip\nobreak\noindent\ignorespaces}

\def\appendix#1#2{\global\meqno=1\global\subsecno=0\global\subsubsecno=0\xdef\secsym{\hbox{#1.}}%
\bigbreak\bigskip\noindent{\bf Appendix \hyperdef\hypernoname{appendix}{#1}%
{#1.} #2}{\toks0{(#1. #2)}\message{\the\toks0}}%
\xdef\s@csym{#1.}\xdef\secn@m{#1}%
\writetoca{{\string\hyperref{}{appendix}{#1}{\bf {#1}\quad}} {\bf #2}}%
\par\nobreak\medskip\nobreak}

%
\def\checkm@de#1#2{\ifmmode{\def\f@rst##1{##1}\hyperdef\hypernoname{equation}%
{#1}{#2}}\else\hyperref{}{equation}{#1}{#2}\fi}
\def\eqnn#1{\DefWarn#1\xdef #1{(\noexpand\relax\noexpand\checkm@de%
{\s@csym\the\meqno}{\secsym\the\meqno})}%
\wrlabeL#1\writedef{#1\leftbracket#1}\global\advance\meqno by1}
\def\f@rst#1{\c@t#1a\em@ark}\def\c@t#1#2\em@ark{#1}
\def\eqna#1{\DefWarn#1\wrlabeL{#1$\{\}$}%
\xdef #1##1{(\noexpand\relax\noexpand\checkm@de%
{\s@csym\the\meqno\noexpand\f@rst{##1}1}{\hbox{$\secsym\the\meqno##1$}})}
\writedef{#1\numbersign1\leftbracket#1{\numbersign1}}\global\advance\meqno by1}
\def\eqn#1#2{\DefWarn#1%
\xdef #1{(\noexpand\hyperref{}{equation}{\s@csym\the\meqno}%
{\secsym\the\meqno})}$$#2\eqno(\hyperdef\hypernoname{equation}%
{\s@csym\the\meqno}{\secsym\the\meqno})\eqlabeL#1$$%
\writedef{#1\leftbracket#1}\global\advance\meqno by1}
\def\xeqn{\expandafter\xe@n}\def\xe@n(#1){#1}
\def\xeqna#1{\expandafter\xe@n#1}
\def\eqns#1{(\e@ns #1{\hbox{}})}
\def\e@ns#1{\ifx\UNd@FiNeD#1\message{eqnlabel \string#1 is undefined.}%
\xdef#1{(?.?)}\fi{\let\hyperref=\relax\xdef\next{#1}}%
\ifx\next\em@rk\def\next{}\else%
\ifx\next#1\xeqn#1\else\def\n@xt{#1}\ifx\n@xt\next#1\else\xeqna#1\fi
\fi\let\next=\e@ns\fi\next}
\def\DefWarn#1{}
%
\newskip\footskip\footskip14pt plus 1pt minus 1pt 
\def\footnotefont{\ninepoint}\def\f@t#1{\footnotefont #1\@foot}
\def\f@@t{\baselineskip\footskip\bgroup\footnotefont\aftergroup\@foot\let\next}
\setbox\strutbox=\hbox{\vrule height9.5pt depth4.5pt width0pt}
\global\newcount\ftno \global\ftno=0
\def\foot{\global\advance\ftno by1\def\foot@rg{\hyperref{}{footnote}%
{\the\ftno}{\the\ftno}\xdef\foot@rg{\noexpand\hyperdef\noexpand\hypernoname%
{footnote}{\the\ftno}{\the\ftno}}}\footnote{$^{\foot@rg}$}}
%
%
%
\global\newcount\refno \global\refno=1
\newwrite\rfile
\def\ref{[\hyperref{}{reference}{\the\refno}{\the\refno}]\nref}
\def\nref#1{\DefWarn#1%
\xdef#1{[\noexpand\hyperref{}{reference}{\the\refno}{\the\refno}]}%
\writedef{#1\leftbracket#1}%
\ifnum\refno=1\immediate\openout\rfile=\jobname.refs\fi
\chardef\wfile=\rfile\immediate\write\rfile{\noexpand\item{[\noexpand\hyperdef%
\noexpand\hypernoname{reference}{\the\refno}{\the\refno}]\ }%
\reflabeL{#1\hskip.31in}\pctsign}\global\advance\refno by1\findarg}
\def\findarg#1#{\begingroup\obeylines\newlinechar=`\^^M\pass@rg}
{\obeylines\gdef\pass@rg#1{\writ@line\relax #1^^M\hbox{}^^M}%
\gdef\writ@line#1^^M{\expandafter\toks0\expandafter{\striprel@x #1}%
\edef\next{\the\toks0}\ifx\next\em@rk\let\next=\endgroup\else\ifx\next\empty%
\else\immediate\write\wfile{\the\toks0}\fi\let\next=\writ@line\fi\next\relax}}
\def\striprel@x#1{} \def\em@rk{\hbox{}}
\def\lref{\begingroup\obeylines\lr@f}
\def\lr@f#1#2{\DefWarn#1\gdef#1{\let#1=\UNd@FiNeD\ref#1{#2}}\endgroup\unskip}
\def\semi{;\hfil\break}
\def\addref#1{\immediate\write\rfile{\noexpand\item{}#1}} 
\def\listrefs{\vfill\supereject\immediate\closeout\rfile\writestoppt
\baselineskip=\footskip\centerline{{\bf References}}\bigskip{\parindent=20pt%
\frenchspacing\escapechar=` \input \jobname.refs\vfill\eject}\nonfrenchspacing}
\def\startrefs#1{\immediate\openout\rfile=\jobname.refs\refno=#1}
\def\xref{\expandafter\xr@f}\def\xr@f[#1]{#1}
\def\refs#1{\count255=1[\r@fs #1{\hbox{}}]}
\def\r@fs#1{\ifx\UNd@FiNeD#1\message{reflabel \string#1 is undefined.}%
\nref#1{need to supply reference \string#1.}\fi%
\vphantom{\hphantom{#1}}{\let\hyperref=\relax\xdef\next{#1}}%
\ifx\next\em@rk\def\next{}%
\else\ifx\next#1\ifodd\count255\relax\xref#1\count255=0\fi%
\else#1\count255=1\fi\let\next=\r@fs\fi\next}
%

%
\newwrite\ffile\global\newcount\figno \global\figno=1
\def\fig{fig.~\hyperref{}{figure}{\the\figno}{\the\figno}\nfig}
\def\nfig#1{\DefWarn#1%
\xdef#1{fig.~\noexpand\hyperref{}{figure}{\the\figno}{\the\figno}}%
\writedef{#1\leftbracket fig.\noexpand~\xfig#1}%
\ifnum\figno=1\immediate\openout\ffile=\jobname.figs\fi\chardef\wfile=\ffile%
{\let\hyperref=\relax
\immediate\write\ffile{\noexpand\medskip\noexpand\item{Fig.\ %
\noexpand\hyperdef\noexpand\hypernoname{figure}{\the\figno}{\the\figno}. }
\reflabeL{#1\hskip.55in}\pctsign}}\global\advance\figno by1\findarg}
\def\xfig{\expandafter\xf@g}\def\xf@g fig.\penalty\@M\ {}
\def\figs#1{figs.~\f@gs #1{\hbox{}}}
\def\f@gs#1{{\let\hyperref=\relax\xdef\next{#1}}\ifx\next\em@rk\def\next{}\else
\ifx\next#1\xfig #1\else#1\fi\let\next=\f@gs\fi\next}
%
\def\figin{\epsfcheck\figin}\def\figins{\epsfcheck\figins}
\def\epsfcheck{\ifx\epsfbox\UnDeFiNeD
\message{(NO epsf.tex, FIGURES WILL BE IGNORED)}
\gdef\figin##1{\vskip2in}\gdef\figins##1{\hskip.5in}
\else\message{(FIGURES WILL BE INCLUDED)}%
\gdef\figin##1{##1}\gdef\figins##1{##1}\fi}
\def\figinsert{\goodbreak\topinsert}
\def\ifig#1#2#3{\DefWarn#1\xdef#1{fig.~\the\figno}
\writedef{#1\leftbracket fig.\noexpand~\the\figno}%
\figinsert\figin{\centerline{#3}}
\smallskip
\leftskip=0pt \rightskip=0pt
\baselineskip12pt\noindent
{{\bf Fig.~\the\figno}\ \ninepoint #2}
\medskip
\global\advance\figno by1\par\endinsert}
\newwrite\lfile
{\escapechar-1\xdef\pctsign{\string\%}\xdef\leftbracket{\string\{}
\xdef\rightbracket{\string\}}\xdef\numbersign{\string\#}}
\def\writedefs{\immediate\openout\lfile=label.defs \def\writedef##1{%
{\let\hyperref=\relax\let\hyperdef=\relax\let\hypernoname=\relax
 \immediate\write\lfile{\string\checkdef\string##1\rightbracket}}}}%
\def\writestop{\def\writestoppt{\immediate\write\lfile{\string\pageno
 \the\pageno\string\startrefs\leftbracket\the\refno\rightbracket
 \string\def\string\secsym\leftbracket\secsym\rightbracket
 \string\secno\the\secno\string\meqno\the\meqno}\immediate\closeout\lfile}}
\def\writestoppt{}\def\writedef#1{}

\def\seclab#1\par{\DefWarn#1%
\xdef #1{\noexpand\hyperref{}{section}{\the\secno}{\the\secno}}%
\writedef{#1\leftbracket#1}\wrlabeL{#1=#1}\par%
\nobreak\medskip\nobreak\noindent\ignorespaces}
\def\subseclab#1\par{\DefWarn#1%
\xdef #1{\noexpand\hyperref{}{subsection}{\the\secno.\the\subsecno}%
{\the\secno.\the\subsecno}}\writedef{#1\leftbracket#1}\wrlabeL{#1=#1}\par%
\nobreak\medskip\nobreak\noindent\ignorespaces}
\def\subsubseclab#1\par{\DefWarn#1%
\xdef#1{\noexpand\hyperref{}{subsubsection}{\the\secno.\the\subsecno.\the\subsubsecno}%
{\the\secno.\the\subsecno.\the\subsubsecno}}\writedef{#1\leftbracket#1}\wrlabeL{#1=#1}\par%
\nobreak\medskip\nobreak\noindent\ignorespaces}
\def\applab#1\par{\DefWarn#1%
\xdef#1{\noexpand\hyperref{}{appendix}{\secn@m}{\secn@m}}%
\writedef{#1\leftbracket#1}\wrlabeL{#1=#1}%
\par\nobreak\medskip\nobreak\noindent\ignorespaces}
\def\appsublab#1{\DefWarn#1%
\xdef #1{\noexpand\hyperref{}{appendix}{\secn@m.\the\subsecno}{\secn@m.\the\subsecno}}%
\writedef{#1\leftbracket#1}\wrlabeL{#1=#1}}
\newwrite\tfile \def\writetoca#1{}
\def\leaderfill{\leaders\hbox to 1em{\hss.\hss}\hfill}
\def\writetoc{\immediate\openout\tfile=\jobname.toc
   \def\writetoca##1{{\edef\next{\write\tfile{\noindent ##1
   \string\leaderfill{
   \string\hyperref{}{page}{\noexpand\number\pageno}%
   {\noexpand\number\pageno}} \par}}\next}}
}
\newread\ch@ckfile
\def\listtoc{\immediate\closeout\tfile\immediate\openin\ch@ckfile=\jobname.toc
\ifeof\ch@ckfile\message{no file \jobname.toc, no table of contents this pass}%
\else\closein\ch@ckfile\centerline{\bf Contents}\nobreak\medskip%
{\baselineskip=15.5pt\footnotefont\parskip=0pt\catcode`\@=11\input\jobname.toc
\catcode`\@=12\bigbreak\bigskip}\fi}
\catcode`\@=12 
\def\tenpoint{\def\rm{\fam0\tenrm}
\textfont0=\tenrm \scriptfont0=\sevenrm \scriptscriptfont0=\fiverm
\textfont1=\teni  \scriptfont1=\seveni  \scriptscriptfont1=\fivei
\textfont2=\tensy \scriptfont2=\sevensy \scriptscriptfont2=\fivesy
\textfont\itfam=\tenit \def\it{\fam\itfam\tenit}\def\footnotefont{\ninepoint}%
\textfont\bffam=\tenbf \def\bf{\fam\bffam\tenbf}\def\sl{\fam\slfam\tensl}\rm}
\font\ninerm=cmr9 \font\sixrm=cmr6 \font\ninei=cmmi9 \font\sixi=cmmi6
\font\ninesy=cmsy9 \font\sixsy=cmsy6 \font\ninebf=cmbx9
\font\nineit=cmti9 \font\ninesl=cmsl9 \skewchar\ninei='177
\skewchar\sixi='177 \skewchar\ninesy='60 \skewchar\sixsy='60
\def\ninepoint{\def\rm{\fam0\ninerm}
\textfont0=\ninerm \scriptfont0=\sixrm \scriptscriptfont0=\fiverm
\textfont1=\ninei \scriptfont1=\sixi \scriptscriptfont1=\fivei
\textfont2=\ninesy \scriptfont2=\sixsy \scriptscriptfont2=\fivesy
\textfont\itfam=\ninei \def\it{\fam\itfam\nineit}\def\sl{\fam\slfam\ninesl}%
\textfont\bffam=\ninebf \def\bf{\fam\bffam\ninebf}\rm}
%
\hyphenation{anom-aly anom-alies coun-ter-term coun-ter-terms}

\def\tikzcaption#1#2{\DefWarn#1\xdef#1{Fig.~\the\figno}
\writedef{#1\leftbracket Fig.\noexpand~\the\figno}%
{
\smallskip
\leftskip=20pt \rightskip=20pt \baselineskip12pt\noindent
{{\bf Fig.~\the\figno}\ \ninepoint #2}
\bigskip
\global\advance\figno by1 \par}}

\def\ntoalpha#1{%
\ifcase#1%
@%
\or A\or B\or C\or D\or E\or F\or G\or H\or I\or J\or K\or L\or M%
\fi
}

\global\newcount\appno \global\appno=1
\def\applab#1{\xdef #1{\ntoalpha{\appno}}\writedef{#1\leftbracket#1}\wrlabeL{#1=#1}
\global\advance\appno by1}

\def\preprint#1 #2\par{\rightline{\vbox{\baselineskip12pt\hbox{#1}\hbox{#2}}}\vskip2cm}
%
\def\title#1\par{\centerline{\bf #1}\nopagenumbers\pageno=0}
\def\author#1\par{\bigskip\bigskip\centerline{#1}}

\newcount\addressno

\def\email#1#2{
\footnote{\null}{\kern-\parindent \llap{$^#1$\hskip1pt}email: #2}}

\def\startcenter{%
  \par
  \begingroup
  \leftskip=0pt plus 1fil
  \rightskip=\leftskip
  \parindent=0pt
  \parfillskip=0pt
}
\def\stopcenter{\endgroup}

\def\address{\bigskip%
  \ifnum\the\addressno=0\else\stopcenter\endgroup\fi
  \advance\addressno by 1%
  \begingroup
  \startcenter
  \it
  \obeylines
  \addressAux
}
\def\addressAux#1{#1}

\def\abstract{\stopcenter\endgroup\bigskip\bigskip\noindent}

\def\Dsl{\,\raise.15ex\hbox{/}\mkern-13.5mu D} 
\def\dsl{\raise.15ex\hbox{/}\kern-.57em\partial}
 
\def\boxeqn#1{\vcenter{\vbox{\hrule\hbox{\vrule\kern3pt\vbox{\kern3pt
	\hbox{${\displaystyle #1}$}\kern3pt}\kern3pt\vrule}\hrule}}}

\def\lie{\hbox{\it\$}} 

\def\ap{{\alpha^{\prime}}}

\def\a{\alpha}
\def\b{{\beta}}
\def\g{{\gamma}}
\def\d{{\delta}}

\def\l{\lambda}

\def\t{{\theta}}

\def\half{{1\over 2}}
\def\p{{\partial}}

\def\({\left(}
\def\){\right)}

\def\cA{{\cal A}}
\def\cF{{\cal F}}

\def\cK{{\cal K}}

\def\cW{{\cal W}}


\def\bA{{\Bbb A}}
\def\bW{{\Bbb W}}
\def\bF{{\Bbb F}}

\def\Box{\square}
\def\AYM{A^{\rm SYM}}


\def\len#1{{%
\def\Dlen{\left|\mkern-1mu #1\mkern -0.5mu\right|}%
\def\Sslen{\left|\mkern-1.3mu #1\mkern -1.3mu\right|}%
\def\SSlen{\left|\mkern-2.8mu #1\mkern-1.3mu\right|}%
\mathchoice{\Dlen}{\Dlen}{\Sslen}{\SSlen}}}

\def\sfrac#1/#2{\kern.1em\raise.5ex\hbox{\the\scriptfont0 #1}%
\kern-.1em/\kern-.15em\lower.25ex\hbox{\the\scriptfont0 #2}}

\font\tenshuffle=shuffle10 \font\sevenshuffle=shuffle7 \font\fiveshuffle=shuffle7 at 5pt
\def\shuffle{{%
\def\Dshuffle{\mathbin{\hbox{\tenshuffle\char'001}}}%
\def\Sshuffle{\mathbin{\hbox{\sevenshuffle\char'001}}}%
\def\SSshuffle{\mathbin{\hbox{\fiveshuffle\char'001}}}%
\mathchoice{\Dshuffle}{\Dshuffle}{\Sshuffle}{\SSshuffle}}}


\def\qed{\hbox{\hskip 3pt
\vbox{\hrule\hbox to 7pt{\vrule height 7pt\hfill\vrule}
\hrule}}\hskip3pt}

\overfullrule=0pt\relax

\frenchspacing

\def\checkdef#1#2{
\ifx\UndeFined#1%
	\def#1{#2}
\else
	\immediate\write16{*** BUG ***: the label \string#1 is already defined ***}
\fi
}
\newread\instream

\openin\instream= label.defs
\ifeof\instream\message{No labels in advance yet. Wait till next pass.}
\else\closein\instream \input label.defs
\fi
\writedefs

\def\arXiv:#1].{\hepthStrip#1 \nil}
\def\hepthStrip#1 #2\nil{\href{http://arxiv.org/abs/#1}{arXiv:#1 #2\unskip}].}


\input epsf.tex

\def\frac#1#2{{#1\over #2}}
\def\Hhat{{\hat H}}
\def\Ahat{{\hat A}}
\def\What{{\hat W}}
\def\Fhat{{\hat F}}
\def\cH{{\cal H}}
\def\dlb{{\lbrack\!\lbrack}}
\def\drb{{\rbrack\!\rbrack}}
\def\lnabla{\nabla^{(L)}} 
\def\twedge{\mathop{\tilde\wedge}}

\title Algorithmic construction of SYM multiparticle superfields in the BCJ gauge

\author
Elliot Bridges\email{\dagger}{e.n.bridges@soton.ac.uk}$^{\dagger}$ and
Carlos R. Mafra\email{\ddagger}{c.r.mafra@soton.ac.uk}$^{\ddagger}$

\address
Mathematical Sciences and STAG Research Centre, University of Southampton,
Highfield, Southampton, SO17 1BJ, UK

\abstract
We write down closed formulas for all necessary steps to obtain
multiparticle super Yang--Mills superfields in the so-called BCJ gauge. The superfields in
this gauge have obvious applications in the quest for finding BCJ-satisfying
representations of amplitudes. As a benefit of having these closed formulas,
we identify the explicit {\it finite} gauge transformation responsible for
attaining the BCJ gauge. To do this, several combinatorial maps on words are
introduced and associated identities rigorously proven.

\Date{June 2019}


\lref\kinalJ{
	G.~Chen, H.~Johansson, F.~Teng and T.~Wang,
  	``On the kinematic algebra for BCJ numerators beyond the MHV sector,''
	[arXiv:1906.10683 [hep-th]].
}

\lref\kinalMon{
	R.~Monteiro and D.~O'Connell,
  	``The Kinematic Algebra From the Self-Dual Sector,''
	JHEP {\bf 1107}, 007 (2011).
	[arXiv:1105.2565 [hep-th]].
}

\lref\BGFthree{
	L.M.~Garozzo, L.~Queimada and O.~Schlotterer,
  	``Berends-Giele currents in Bern-Carrasco-Johansson gauge for $F^3$- and $F^4$-deformed Yang-Mills amplitudes,''
	JHEP {\bf 1902}, 078 (2019).
	[arXiv:1809.08103 [hep-th]].
}

\lref\linftyW{
	T.~Macrelli, C.~Sämann and M.~Wolf,
  	``Scattering Amplitude Recursion Relations in BV Quantisable Theories,''
	[arXiv:1903.05713 [hep-th]].
}

\lref\blessenohl{
	Blessenohl, D. and Laue, H., ``Generalized Jacobi identities,'' Note di
	Matematica 8, no. 1 (1988) 111-121.
}

\lref\BGSym{
	F.A.~Berends and W.T.~Giele,
	``Multiple Soft Gluon Radiation in Parton Processes,''
	Nucl.\ Phys.\ B {\bf 313}, 595 (1989).
}

\lref\thibon{
J.-Y. Thibon,
``Lie idempotents in descent algebras''
(lecture notes), Workshop on Hopf Algebras and Props, Boston, March 5 - 9, 2007 Clay Mathematics Institute.
}

\lref\exPSS{
	N.~Berkovits,
  	``Explaining Pure Spinor Superspace,''
	[hep-th/0612021].
}

\lref\Newnathan{
	N.~Berkovits,
	``Untwisting the pure spinor formalism to the RNS and twistor string in a flat and AdS$_{5} \times$ S$^{5}$ background,''
	JHEP {\bf 1606}, 127 (2016).
	[arXiv:1604.04617 [hep-th]].
}

\lref\BGPS{
	C.R.~Mafra and O.~Schlotterer,
  	``Berends-Giele recursions and the BCJ duality in superspace and components,''
	JHEP {\bf 1603}, 097 (2016).
	[arXiv:1510.08846 [hep-th]].
}
\lref\psf{
 	N.~Berkovits,
	``Super-Poincare covariant quantization of the superstring,''
	JHEP {\bf 0004}, 018 (2000)
	[arXiv:hep-th/0001035].
	\semi
	N.~Berkovits,
  	``ICTP lectures on covariant quantization of the superstring,''
	ICTP Lect.\ Notes Ser.\  {\bf 13}, 57 (2003).
	[hep-th/0209059].
}
\lref\BrinkBC{
	L.~Brink, J.H.~Schwarz and J.~Scherk,
  	``Supersymmetric Yang-Mills Theories,''
	Nucl.\ Phys.\ B {\bf 121}, 77 (1977)..
}

\lref\selivanov{
	K.G.~Selivanov,
  	``On tree form-factors in (supersymmetric) Yang-Mills theory,''
	Commun.\ Math.\ Phys.\  {\bf 208}, 671 (2000).
	[hep-th/9809046].
	\semi
	K.G.~Selivanov,
  	``Postclassicism in tree amplitudes,''
	[hep-th/9905128].
}

\lref\EOMbbs{
	C.R.~Mafra and O.~Schlotterer,
  	``Multiparticle SYM equations of motion and pure spinor BRST blocks,''
	JHEP {\bf 1407}, 153 (2014).
	[arXiv:1404.4986 [hep-th]].
}
\lref\Gauge{
	S.~Lee, C.R.~Mafra and O.~Schlotterer,
  	``Non-linear gauge transformations in $D=10$ SYM theory and the BCJ duality,''
	JHEP {\bf 1603}, 090 (2016).
	[arXiv:1510.08843 [hep-th]].
}
\lref\nptMethod{
	C.R.~Mafra, O.~Schlotterer, S.~Stieberger and D.~Tsimpis,
	``A recursive method for SYM n-point tree amplitudes,''
	Phys.\ Rev.\ D {\bf 83}, 126012 (2011).
	[arXiv:1012.3981 [hep-th]].
}
\lref\oneloopI{
	C.R.~Mafra and O.~Schlotterer,
  	``Towards the n-point one-loop superstring amplitude I: Pure spinors and superfield kinematics,''
	[arXiv:1812.10969 [hep-th]].
}

\lref\Reutenauer{
	C.~Reutenauer,
	``Free Lie Algebras,''
	London Mathematical Society Monographs, 1993
}

\lref\NptString{
	C.R.~Mafra, O.~Schlotterer and S.~Stieberger,
  	``Complete N-Point Superstring Disk Amplitude I. Pure Spinor Computation,''
	Nucl.\ Phys.\ B {\bf 873}, 419 (2013).
	[arXiv:1106.2645 [hep-th]].
}

\lref\NonLinEquations{
	C.~R.~Mafra and O.~Schlotterer,
	``Solution to the nonlinear field equations of ten dimensional supersymmetric Yang-Mills theory,''
	Phys.\ Rev.\ D {\bf 92}, no. 6, 066001 (2015).
	[arXiv:1501.05562 [hep-th]].
}
\lref\wittentwistor{
	E.Witten,
        ``Twistor-Like Transform In Ten-Dimensions''
        Nucl.Phys. B {\bf 266}, 245~(1986).
}
\lref\SiegelYI{
	W.~Siegel,
  	``Superfields in Higher Dimensional Space-time,''
	Phys.\ Lett.\ B {\bf 80}, 220 (1979).
}
\lref\barcelo{
	Barcelo, H. and Sundaram, S., ``On Some Submodules of the Action of the Symmetrical
	Group on the Free Lie Algebra''. Journal of Algebra, 154(1), (1993) pp.12-26.
}
\lref\PScomb{
	C.R.~Mafra, ``Planar binary trees in scattering amplitudes'', CARMA
	2017 proceedings.
}
\lref\FORM{
	J.A.M.~Vermaseren,
	``New features of FORM,''
	arXiv:math-ph/0010025.
\semi
	M.~Tentyukov and J.A.M.~Vermaseren,
	``The multithreaded version of FORM,''
	arXiv:hep-ph/0702279.
}
\lref\PSBCJ{
	C.R.~Mafra, O.~Schlotterer and S.~Stieberger,
	``Explicit BCJ Numerators from Pure Spinors,''
	JHEP {\bf 1107}, 092 (2011).
	[arXiv:1104.5224 [hep-th]].
}
\lref\BCJ{
	Z.~Bern, J.J.M.~Carrasco and H.~Johansson,
	``New Relations for Gauge-Theory Amplitudes,''
	Phys.\ Rev.\ D {\bf 78}, 085011 (2008).
	[arXiv:0805.3993 [hep-ph]].
	\semi
  	Z.~Bern, J.J.M.~Carrasco and H.~Johansson,
  	``Perturbative Quantum Gravity as a Double Copy of Gauge Theory,''
	Phys.\ Rev.\ Lett.\  {\bf 105}, 061602 (2010).
	[arXiv:1004.0476 [hep-th]].
}

\lref\BGpaper{
	F.A.~Berends, W.T.~Giele,
  	``Recursive Calculations for Processes with n Gluons,''
	Nucl.\ Phys.\  {\bf B306}, 759 (1988).
}
\lref\nptFT{
	C.R.~Mafra, O.~Schlotterer, S.~Stieberger and D.~Tsimpis,
	``A recursive method for SYM n-point tree amplitudes,''
	Phys.\ Rev.\ D {\bf 83}, 126012 (2011).
	[arXiv:1012.3981 [hep-th]].
}
\lref\nptString{
	C.R.~Mafra, O.~Schlotterer and S.~Stieberger,
  	``Complete N-Point Superstring Disk Amplitude I. Pure Spinor Computation,''
	Nucl.\ Phys.\ B {\bf 873}, 419 (2013).
	[arXiv:1106.2645 [hep-th]].
}
\lref\towardsTwo{
	C.R.~Mafra and O.~Schlotterer,
  	``Two-loop five-point amplitudes of super Yang-Mills and supergravity in pure spinor superspace,''
	JHEP {\bf 1510}, 124 (2015).
	[arXiv:1505.02746 [hep-th]].
}
\lref\towardsOne{
	C.R.~Mafra and O.~Schlotterer,
  	``Towards one-loop SYM amplitudes from the pure spinor BRST cohomology,''
	Fortsch.\ Phys.\  {\bf 63}, no. 2, 105 (2015).
	[arXiv:1410.0668 [hep-th]].
}

\listtoc
\writetoc
\filbreak

\newsec Introduction

The definition and usage of multiparticle superfields \refs{\EOMbbs,\Gauge} of
supersymmetric Yang--Mills (SYM) theory \BrinkBC\ has proved to be an
essential feature in obtaining compact expressions for high-multiplicity
amplitudes in superstring \nptString\ and field theories \nptFT\ using the
pure spinor formalism \psf.

In the simplest formulation of multiparticle superfields in the {\it Lorenz
gauge}, their definition is given by a straightforward recursion over the particle labels \Gauge.
While this recursive definition has its own merits and is certainly useful
in relating the new expressions for tree-level amplitudes \BGPS\ to the standard
Berends--Giele recursions \BGpaper, there is an alternative formulation related by
a non-linear gauge transformation whose properties have more appeal,
the {\it BCJ-gauge} representation \EOMbbs. As will be reviewed in
section~\BCJcharacsec, the superfields in this gauge satisfy {\it generalized
Jacobi identities} \blessenohl\ in their particle labels, for example $A^m_{12}=-A^m_{21}$, 
$A^m_{123}+A^m_{231}+A^m_{312}=0$, and so forth. In this gauge, they constitute
the
natural building blocks used in the expressions of local SYM numerators
satisfying the Bern--Carrasco--Johansson numerator identities \BCJ\
at tree- \PSBCJ\ and loop-level \refs{\towardsOne,\towardsTwo}.

As explained in \Gauge, the gauge transformations required to go to the BCJ
gauge are encoded in so-called {\it redefining} superfields $H_{[P,Q]}$ to be
reviewed below. Until now, the explicit expressions of these superfields were
known only up to multiplicity five \Gauge. In section~\HPQsols\ of this paper
this restriction will be lifted when we propose a recursive formula
for $H_{[P,Q]}$, namely
\eqnn\genHintro
$$\eqalignno{
H_{[P,Q]} &=
(-1)^{\len{Q}}{\len{P}\over\len{P}+\len{Q}}
\sum_{XjY=\dot p\tilde Q}(-1)^{|Y|}H'_{\tilde Y,j,X} - (P\leftrightarrow
Q)\,,\qquad H_{[i,j]}=0\,,&\genHintro\cr
}$$
where the auxiliary superfields $H'_{A,B,C}$ are defined by
\eqnn\auxH
$$\eqalignno{
H'_{P,Q,R} &\equiv H_{P,Q,R} +\Big[\half H_{[P,Q]}(k_{PQ}\cdot A_R)  + {\rm
cyclic}(P,Q,R)\Big]\cr
& -\Big[\!\!\!\! \sum_{XjY=P\atop \delta(Y)=R\otimes S}
\!\!\!\!\!(k^{X}\cdot k^j)\big[H_{[XR,Q]}H_{[jS,R]} -(X\leftrightarrow j)\big]
 + {\rm cyclic}(P,Q,R)\Big]\,,\cr
H_{P,Q,R} &\equiv -{1\over 4}A^m_P A^n_Q F^{mn}_R
+ \half (W_P\g_m W_Q)A_R^m + {\rm cyclic}(P,Q,R)\,.
}$$
As a consequence of the quadratic corrections $H^2$ in these formulas, we will
show in section~\gaugetranssec\ that the superfields satisfying the generalized Jacobi identities
follow from a {\it standard} gauge transformation of SYM theory in its
{\it finite} form,
\eqn\finitegauIntro{
\Bbb A_m^{\rm BCJ} = U\Bbb A_m^{\rm L} U^{-1} + \p_mU U^{-1}\hbox{ with } U =
\exp(-\Bbb H)\,,
}
whose series representation is given by
\eqn\finiteIntro{
\Bbb A_m^{{\rm BCJ}} =
\Bbb A_m^{{\rm L}}
+ [\Bbb H,\p_m] - [\Bbb H, \Bbb A_m^{{\rm L}}]
- \half [\Bbb H, [\Bbb H,\p_m]] + \half [\Bbb H,[\Bbb H, \Bbb A^{\rm L}_m]]
+ {1\over 3!} [\Bbb H,[\Bbb H, [\Bbb H,\p_m]]]
+ \cdots
}
We note that in \Gauge\ only the first three terms of \finiteIntro\ were
identified.

While in pursuit of finding these formulas we also filled some gaps of the
previous discussions. These mostly concern writing down closed formulas for
expressing contact terms (in a multitude of different situations) where the
multiparticle labels are given in terms of an arbitrary configuration of
nested Lie brackets. As will be explained in section~\contactapp, we found a
novel recursive description of such terms which is {\it universal} and whose
backbone is given by the solution to a purely combinatorial problem. Several
equations relevant to the framework of multiparticle superfields can be
written down using this newly found recursion and we prove several associated
results.

Finally, in the appendices we write down some longer examples of applications
of several recursive maps from the main text, among other things.

\newsec Review

In this section we review some aspects of the construction of 10d
supersymmetric Yang--Mills superfields following the recent discussions of
\refs{\Gauge,\EOMbbs} using the framework of perturbiners \selivanov. For the
original references on the covariant description of super Yang--Mills in ten
dimensions, see \refs{\wittentwistor,\SiegelYI}

\newsubsec\notationsec Notation and conventions

\newsubsubsec\tendconv Ten-dimensional superspace

The ten-dimensional superspace coordinates are denoted $\{ x^m,\t^\a\}$, where
$m=0, \ldots,9$ are the vector indices and $\a=1, \ldots,16$ denote the spinor
indices of the Lorentz group. The spinor representation is based on the
$16\times 16$ Pauli matrices $\g^m_{\a\b}=\g^m_{\b\a}$ satisfying the Clifford
algebra $\g^{(m}_{\a\b}\g^{n)\b\g}_{\phantom\a} = 2\eta^{mn}\d_\a^\g$. In this
paper the (anti)symmetrization of $n$ indices does not include a factor of
${1\over n!}$.

\newsubsubsec\notationsec Multiparticle index notation

In the following discussions we will use a notation based on ``words''
composed of ``letters'' from the alphabet of natural numbers. Capital letters
from the Latin alphabet are used to represent words (e.g. $P=1423$) while
their composing letters are represented by lower case letters (e.g. $i=3$).
The length of a word $P$ is denoted $\len{P}$ and it is given by the number of
its letters. The reversal of a word $P=p_1p_2 \ldots p_\len{P}$ is $\tilde
P=p_\len{P} \ldots p_2 p_1$. The word notation is also used in place of
arbitrary commutators, such as $P=[1,2]\equiv 12 - 21$; the context will
disambiguate whether a word denotes a sequence of letters or a bracketing
structure. In addition, when the bracketing structure is nested from left to
right such as $P=[[[[1,2],3],4],5]$ we will often write it as $P=12345$. Such
structures may be referred to as (left-to-right) ``Dynkin brackets''

The multiparticle momentum for a word with letters (labels) from massless particles
$(k_i\cdot k_i)=0$ and its associated Mandelstam invariant are given by
\eqn\mandef{
k^m_P\equiv k^m_{p_1}+ \cdots +k^m_{p_\len{P}}\,,\qquad s_P \equiv \half (k_P\cdot k_P)\,.
}
For example $k^m_{123}\equiv k^m_1 + k^m_2 + k^m_3$ and $s_{123} = s_{12}+s_{13}+s_{23}$.

\newsubsec\nonlinSYMsec Non-linear supersymmetric Yang--Mills

To describe ten-dimensional SYM one introduces Lie algebra-valued superfield connections $\Bbb
A_\a = \Bbb A_\alpha(x,\t)$ and $\Bbb A_m = \Bbb A_m(x,\t)$ and the
supercovariant derivatives \refs{\SiegelYI, \wittentwistor},
\eqn\covder{
\nabla_\a \equiv D_\a - \Bbb A_\a\,,\qquad \nabla_m \equiv \p_m - \Bbb A_m\,,
}
where the superspace derivative $D_\a \equiv {\p\over\p\t^\a} + {1\over 2}(\g^m\t)_\a\p_m$
satisfies $\{D_\a,D_\b\} = \g^m_{\a\b}\p_m$.
The constraint
$\big\{ \nabla_\alpha ,\nabla_\beta \big\} = \gamma^m_{\alpha \beta} \nabla_m$
and the associated Bianchi identities imply the following
non-linear equations of motion \wittentwistor,
\eqn\SYMeomO{
\eqalign{
\big\{ \nabla_\a , \nabla_\b \big\} &= \g^m_{\a\b} \nabla_m\,, \cr
\big[ \nabla_\a , \nabla_m\big] &= - (\g_m \bW)_\a\,,
}
\qquad\eqalign{
\big\{ \nabla_{\a} ,\bW^\beta \big\} &= {1\over 4}
(\gamma^{mn})_{\a}{}^{\beta} \bF_{mn}\,, \cr
\big[\nabla_{\a}, \bF^{mn} \big] &=  (\Bbb W^{[m} \gamma^{n]})_\a \,,
}}
where
\eqn\WWWdef{
\Bbb F_{mn} \equiv - \big[  \nabla_m, \nabla_n \big] \ , \qquad
\Bbb W^{\a}_m \equiv \big[  \nabla_m, \Bbb W^\a \big]
\,.
}
These equations are invariant under the gauge transformations of
the superpotentials
\eqnn\NLgauge
$$\eqalignno{
\delta_{\Omega } \Bbb A_\a & = \big[\nabla_\a ,\Omega \big]\,,
\qquad \delta_\Omega \Bbb A_m = \big[\nabla_m ,\Omega\big] &\NLgauge
}
$$
which in turn induce the gauge transformations of their field-strengths
$\delta_\Omega\Bbb W^\a =  \big[\Omega, \Bbb W^\a  \big]$,
$\delta_{\Omega }  \Bbb F^{mn} =  \big[\Omega, \Bbb F^{mn}  \big]$, and
$\delta_{\Omega }  \Bbb W^\a_m =  \big[\Omega, \Bbb W^\a_m  \big]$
where $\Omega\equiv\Omega(x,\theta)$ is a Lie algebra-valued gauge parameter superfield.
The equations of motion \SYMeomO\ can also be rewritten as
\eqn\SYMeom{
\eqalign{
\{\nabla_{(\a},\Bbb A_{\b) } \}&=\gamma^{m}_{\a\b}\Bbb A_m\cr
[\nabla_{\a},\Bbb A_m]&=[\p_m,\Bbb A_\a]+(\gamma_m \Bbb W)_\a\,,
}\quad\eqalign{
\{\nabla_\a,\Bbb W^\b\}&={1\over4}(\g^{mn})_\a^{\phantom{\a}\beta}\Bbb F_{mn}\cr
[\nabla_\a, \Bbb F^{mn}]&= (\Bbb W^{[m} \g^{n]})_\a\,.
}}

\newsubsubsec\nonwavesec Non-linear wave equations and Berends--Giele supercurrents

Alternatively, in the {\it Lorenz gauge} (defined by the constraint $[\p_m,\bA^m]=0$),
the equations of motion \SYMeomO\ are equivalent to the non-linear wave equations \Gauge,
\eqnn\nonwave
$$\eqalignno{
\Box \Bbb A_{\a} &=
\big[ \Bbb A_m ,[\p^m ,\Bbb A_\a]\big] + \big[ (\gamma^m \Bbb
W)_\a, \Bbb A_m \big] &\nonwave\cr
\Box \Bbb A_m &=
 \big[ \Bbb A_p , [\p^p, \Bbb A^m]  \big]  + \big[ \Bbb F^{mp}, \Bbb A_p \big]
 +\gamma^m_{\a \beta} \{ \Bbb W^\a, \Bbb W^\beta \}\cr
\Box \Bbb W^{\a} &=\big[ \Bbb A_m, [ \p^m, \Bbb W^\a ]  \big] + \big[ \Bbb A^m, \Bbb W^\a_m  \big]
+ {1\over 2} \big[\Bbb F_{mn}, (\gamma^{mn} \Bbb W)^\a \big] \cr
\Box \Bbb F^{mn} &= [\Bbb A_p, [\p^p ,\Bbb F^{mn}]] + [\Bbb A_p,  \Bbb F^{p|mn}]
+ 2 [\Bbb F^{mp}, \Bbb F_p{}^n ] + 4 \{ (\Bbb W^{[m} \g^{n]})_\a , \Bbb
W^\a \}\,,
}$$
where $\Box\Bbb K\equiv[\p^m,[\p_m,\Bbb K]]$ and $\Bbb F^{p|mn}\equiv [\nabla^p,\Bbb F^{mn}]$.

To solve the wave equations \nonwave\ we use the perturbiner method of
Selivanov \selivanov. In this approach, one expands the superfields
$\Bbb K\in\{\Bbb A_\alpha,\Bbb A^m, \Bbb W^\alpha, \Bbb F^{mn} \}$
as a series with respect to the generators $t^{i_j}$ of a Lie algebra summed
over all possible non-empty words $P$ as
\eqn\series{
\Bbb K  \equiv
\sum_P \cK_P t^P\,,\qquad t^P\equiv t^{p_1}t^{p_2}\cdots t^{p_{\len{P}}}\,.
}
After plugging these series in \nonwave\ one learns that
the expansion coefficients $\cK_P\in\{\cA^P_\alpha,\cA^m_P,
\cW^\alpha_P, \cF^{mn}_P \}$
turn out to be the Berends--Giele currents,
\eqn\BGdef{
\cK_P = {1\over s_{P}}\sum_{XY=P}\cK_{[X,Y]}\,,
}
where $s_P= \half k^2_P$ arises from the $\Box$ operator acting on
plane waves of momentum $k^m_P$ and
\eqnn\cAalpha
$$\eqalignno{
\cA^{[P,Q]}_\a &= - \half\bigl[  \cA^{P}_\a (k^{P}\cdot  \cA^Q)
+  \cA^{P}_m (\g^m \cW^Q)_\a - (P\leftrightarrow Q)\bigr]\,,  &\cAalpha\cr
\cA^{[P,Q]}_m &= - \half\bigl[ \cA^{P}_m (k^P\cdot\cA^{Q}) + \cA^{P}_n\cF^Q_{mn}
- (\cW^{P}\g_m \cW^Q) - (P \leftrightarrow Q)\bigr]\,,\cr
\cW_{[P,Q]}^\a &= - \half \bigl[  \cW_{P}^\a (k_P\cdot \cA_Q) + \cW_P^{m\a} \cA^m_Q
+ {1\over 2}(\g_{rs} \cW_P)^\a  \cF_{Q}^{rs}
- (P \leftrightarrow Q) \bigr]\,,\cr
\cF^{mn}_{[P,Q]} &= - {1\over 2} \big[ \cF^{mn}_P (k_P \cdot {\cal A}_Q)
+ \cF_{P}^{p|mn} \cA_p^Q + 2 \cF_P^{mp} \cF_{Q \, p}^n
+ 4 \g^{[m}_{\a\b} \cW_P^{n]\a} {\cal W}_Q^\beta
- (P \leftrightarrow Q) \big]\,.
}$$
Notice that the above Berends--Giele currents are non-local superfields as
they contain inverse factors of Mandelstams variables.

\newsubsubsec\singleSYM Linearized description of 10d SYM

The {\it linearized} description of ten-dimensional
super-Yang--Mills is obtained by discarding the quadratic terms from the
equations of motion \SYMeom\ and yields
\eqn\RankOneEOM{
\eqalign{
D_{\a} A^i_{\b} + D_{\b} A^i_{\a} & = \g^m_{\a\b} A^i_m\,,\cr
D_\a A^i_m &= (\g_m W_i)_\a + \p_m A^i_\a\,,
}\qquad\eqalign{
D_\a F^i_{mn} & = \p_{m} (\g_{n} W_i)_\a - \p_{n} (\g_{m} W_i)_\a\,,\cr
D_\a W_i^{\b} &= {1\over 4}(\g^{mn})^{\phantom{m}\b}_\a F^i_{mn}\,.
}}
In the context of scattering amplitudes, the superfields are labelled with
a distinct natural number $i$ to associate them with
the $i$-th particle taking part in the scattering
process. This association will be generalized below.

\newsubsec\BCJcharacsec Generalized Jacobi identities

As we will discuss below in the context of multiparticle superfields, there is
the notion of a superfield satisfying certain symmetries dubbed {\it BCJ
symmetries} in \Gauge. These symmetries can be given a precise mathematical
characterization in terms of what is called {\it generalized Jacobi
identities} in the mathematics literature \refs{\blessenohl,\Reutenauer}.

Let $A$ be a word and $\ell(A)$ its left-to-right bracketing defined in \lrbracket.
The {\it generalized Jacobi identities} correspond to the elements in the kernel of
$\ell$. For example
\eqn\genjacex{
\ell(12+21) = 0\,,\qquad \ell(123+231+312) = 0\,,
}
which correspond with the antisymmetry and Jacobi identity of the Lie bracket.

Using the identity $\ell(P\ell(Q)) = [\ell(P),\ell(Q)]$ it is easy to see that
$\ell(A\ell(B) + B\ell(A)) = 0$ for any words $A$ and $B$. In addition, due to
the recursive definition of $\ell$ if $\ell(P)=0$ it also follows that
$\ell(PQ)=0$ for any word $Q$. Therefore, for objects labelled by words, the
generalized Jacobi identities can be characterized by an abstract operator
$\lie_k$
\eqn\JacobiIdentities{
\lie_k\circ K_{ABC} \equiv K_{A\ell (B)C}+K_{B\ell (A)C},\quad\hbox{$\forall A,B\neq
\emptyset$ and  $\forall C$ such that $\len{A}+\len{B}=k$.}
}
We emphasize the arbitrary partition of non-empty words $A$ and $B$ in the above
definition (while $C$ can be empty), leading to a non-unique operator $\lie$. For instance
\eqnn\nonuni
$$\eqalignno{
\lie_3\circ K_{123}&=K_{123}-K_{132}+K_{231}\,,\quad\hbox{for $A=1$, $B=23$
and $C=\emptyset$}&\nonuni\cr
\lie_3\circ K_{123}&=K_{123}+K_{312}-K_{321}\,,\quad\hbox{for $A=12$, $B=3$
and $C=\emptyset$}\,.
}$$
Note that if $\lie_2\circ K_{123}=0$ then the right-hand side of the expressions in \nonuni\ agree
and can be written as the cyclic sum $K_{123}+K_{231}+K_{312}$.
\proclaim Definition 1.
The objects $K_P$ are said to satisfy generalized Jacobi identities
iff
\eqn\BCJcond{
\lie_k\circ K_P = 0\,,\quad\forall k\le\len{P}\,.
}
The generalized Jacobi identities are also called BCJ symmetries.
\par
\noindent The defining identities for objects $K_P$ of increasing multiplicities can be
written as
\eqnn\RankThreeJacobi
$$\eqalignno{
K_{12C}+K_{21C}&=0,\quad\forall \;C, &\RankThreeJacobi\cr
K_{123C}+K_{231C}+K_{312C}&=0,\quad\forall\; C,\cr
K_{1234C}+K_{2143C}+K_{3412C}+K_{4321C}&=0,\quad\forall\; C.\cr
}$$
where we have already used the fact that $K_P$ satisfies the BCJ symmetries 
$\lie_k\circ K_P = 0$ for all $k\le\len{P}$ to simplify the appearance of
the above. This fact in general can be used to show the equivalence of the BCJ symmetries for
the various partitions of $P=ABC$ as mentioned after the example \nonuni.

It is not hard to be convinced that the BCJ symmetries
are equivalent to the symmetries of a concatenated
string of structure constants,
$K_{12...p}\leftrightarrow f^{12a_2}f^{a_23a_3}f^{a_34a_4}...f^{a_{p-1}pa_p}$.

If $K_P$ satisfies BCJ symmetries then it is convenient to use the
notation $K_{\ell(P)}\equiv K_P$. In particular, this implies that for
superfields in the BCJ gauge we have \oneloopI,
\eqn\Baker{
K_{[P,Q]} = K_{P\ell(Q)}\,.
}
For example, $K_{[12,34]} = K_{1234} - K_{1243}$.
In addition, it follows from the definitions \JacobiIdentities\ and \BCJcond\
that if $K_P$ with $\len{P}=n$ satisfies generalized
Jacobi identities then
\eqn\LieBasis{
K_{AiB} = - K_{i\ell(A)B}\,,\quad A\neq\emptyset\,,\,\forall B\,,
}
which implies that there is an $(n-1)!$ basis of $K_P$.

\newnewsec\contactapp Contact terms for general Lie polynomials

For the purpose of this paper, $P$ is a {\it Lie polynomial} if it is
a linear combination of words written in terms of (nested) Lie
brackets $[x,y]\equiv xy-yx$. For example $P=[[1,2],3]=123-213-312+321$ is a Lie polynomial
while $Q=123$ is not\foot{It may not be
immediately obvious that a given linear combination of words is a Lie polynomial.
For $P=12-21$ this is clear, but it is harder to see that
       $P=1324
       + 1423
       - 1432
       - 2134
       + 2341
       - 3124
       + 3214
       - 3241
       - 4123
       + 4213
       - 4231
       + 4312$
is the Lie polynomial $P=[[[1,2],3],4]+[[[2,3],4],1]$. A theorem by
Dynkin--Specht-Wever states that if $\ell(P)=\len{P}P$ then $P$ is a Lie
polynomial \Reutenauer, and this fact can be used to find the expression
written in terms of nested Lie brackets \thibon.}.

In this section we will introduce mathematical maps acting on words and Lie
polynomials that will play a central role in later discussions about
several aspects of local and non-local multiparticle superfields.

\newsubsec\BGmapsec Planar binary tree map on words

\noindent A nested Lie bracket can be interpreted as a planar binary tree and vice versa
\barcelo. In the context of tree-level scattering amplitudes one can map each
planar binary tree to a product of inverse Mandelstam invariants. For example
the two binary trees with three leaves are mapped to
\medskip
\centerline{{\epsfxsize=0.39\hsize\epsfbox{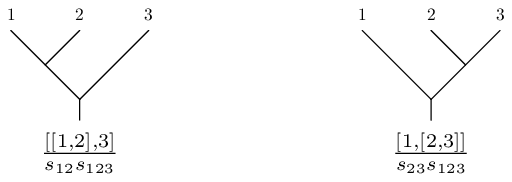}}}
\medskip

\noindent Mapping the sum over all binary trees with a given number of leaves will be
related to Berends--Giele currents later on, and the explicit expansions can
be generated from the following recursion.

\proclaim Definition 2 (Binary tree map). A word $P$ of length $\len{P}$ is recursively mapped to
a Lie polynomial built from a sum over all planar binary trees with $\len{P}$
leaves as
\eqn\BGbr{
b(i) = i,\qquad b(P) = {1\over s_P}\sum_{XY=P}[b(X),b(Y)]\,,
}
where $s_P$ is the Mandelstam invariant \mandef.
\par
\noindent The number of terms in the
recursion above is given by the Catalan numbers
$1,2,5,14, \ldots$ and one gets, for example,
\eqnn\bexamp
$$\displaylines{
b(1) = 1,\quad b(12) = {[1,2]\over s_{12}},
\quad b(123) = {[[1,2],3]\over s_{12}s_{123}}  + {[1,[2,3]]\over
s_{23}s_{123}}\,,\hfil\bexamp\hfilneg\cr
b(1234) = {[ [ [ 1 , 2 ] , 3 ] , 4 ] \over s_{12} s_{123} s_{1234}}
+  {[ [ 1 , [ 2 , 3 ] ] , 4 ] \over s_{123} s_{1234} s_{23}}
+  {[ [ 1 , 2 ] , [ 3 , 4 ] ] \over s_{12} s_{1234} s_{34}}
+  {[ 1 , [ [ 2 , 3 ] , 4 ] ] \over s_{1234} s_{23} s_{234}}
+  {[ 1 , [ 2 , [ 3 , 4 ] ] ] \over s_{1234} s_{234} s_{34}}\,.
}$$
\medskip\noindent
These expansions are easily seen to be examples of {\it Lie
polynomials\/} \Reutenauer, see
figure~\figBGfour\ for the diagrammatic representation of $b(1234)$.

\ifig\figBGfour{The sum generated by the recursion \BGbr\ of $b(1234)$.}
{\epsfxsize=0.88\hsize\epsfbox{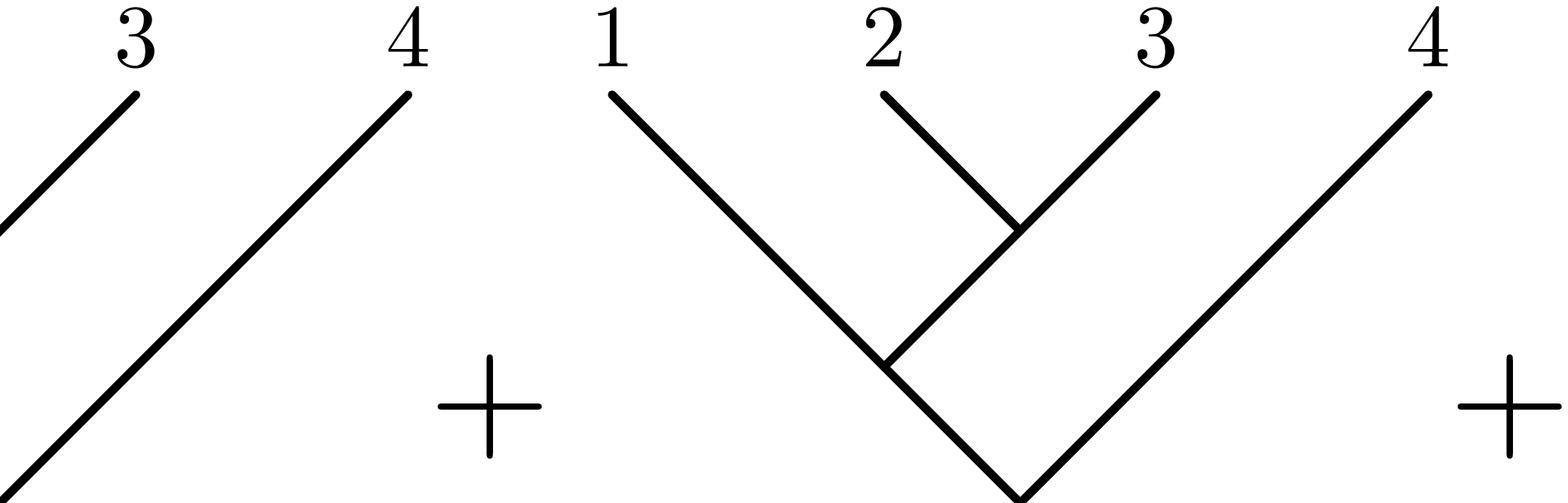}}

\newsubsec\Cmapsec Contact terms associated to Lie polynomials

Given the Lie polynomial $[1,2]$ we can associate to it the following contact terms
proportional to $(k_1\cdot k_2)=s_{12}$;
$C\circ[1,2]\equiv (k_1\cdot k_2)(1\otimes 2 - 2\otimes 1)$.
It is easy to see that this definition leads to a  deconcatenation of $b(12)$,
\eqn\twob{
C\circ b(12) = b(1)\otimes b(2) - b(2)\otimes b(1)
= \sum_{XY=12} \big(b(X)\otimes b(Y) - (X\leftrightarrow Y)\big)\,.
}
We would like to extend this action to an arbitrary Lie polynomial
$C\circ [P,Q]$ such that
\eqn\Cgoal{
C\circ b(P) = \sum_{XY=P}\!\!\! \big(b(X)\otimes b(Y) - (X\leftrightarrow Y)\big)\,.
}
The following definition does the job, as will be proven below.
\proclaim Definition 3 (Contact term map). Let $C$ be the coproduct $C:
{\rm Lie} \rightarrow {\rm Lie}\otimes {\rm Lie}$ that maps a Lie polynomial
into the tensor product of two Lie polynomials recursively by
\eqnn\recurPQ
$$\eqalignno{
C\circ i &\equiv 0 &\recurPQ\cr
C\circ [P,Q] &\equiv \big(C\circ P\big)\wedge Q
+ P\wedge \big(C\circ Q\big)
+ (k_P\cdot k_Q)\big(P\otimes Q - Q\otimes P\big),
}$$
where $\wedge$ is defined by\foot{Note the relations \otdef\ should be used to remove $\wedge$
operations in the reverse order to that which they are introduced. Without
such a criterion ambiguities can arise when objects of the form $A\wedge
[B,C]\wedge D$ are considered.}
\eqnn\otdef
$$\eqalignno{
(A\otimes B)\wedge C &\equiv [A,C]\otimes B + A\otimes [B,C]&\otdef\cr
A\wedge (B\otimes C) &\equiv [A,B]\otimes C + B\otimes [A,C]\,,
}$$
and $k_P^m\equiv k^m_{p_1p_2 \ldots p_{\len{P}}}$, where $p_i$
for $i=1$ to $i=\len{P}$ are the letters of $P$.
\par
\noindent As an immediate consistency check, we note that the definitions
given in \otdef\ imply that $C\circ[Q,P]=-C\circ[P,Q]$.
Note that when the contact term map is used to generate combinations of
superfields, the notation described in \CVTdef\ and \wordRep\ may be used.
For example applications of the $C$ map, see
the appendix~\Cexamplesapp.

\proclaim Proposition 1. The $C$ map satisfies
\eqn\propGoal{
C\circ b(P) = \sum_{XY=P}\!\!\! \big(b(X)\otimes b(Y) - (X\leftrightarrow Y)\big)\,.
}
\par
\noindent{\it Proof.} The proof is inductive in nature. When the word $P$ has
length two the statement has been verified explicitly in \twob.
We now assume that the relation \propGoal\ is satisfied for
any word $P$ of length less than $n$, and let $Q$ be a word of length $n$.
Then we get
\eqnn\propStepOne
$$\displaylines{
s_QC\circ b(Q)=C\circ \sum_{XY=Q}[b(X),b(Y)]\hfil\propStepOne\hfilneg\cr
=\sum_{XY=Q}\!\!\Big[ (C\circ b(X))\wedge b(Y) + b(X)\wedge (C\circ b(Y))
+(k^X\cdot k^Y)(b(X)\otimes b(Y)-b(Y)\otimes b(X))\Big]
}$$
where we have used the definition of the
contact term algorithm \recurPQ.
Now we separate the above into the three possible cases;
both of $\len{X}$ and $\len{Y}$ being greater than $1$, $\len{X}=1$, and
$\len{Y}=1$. We then use that $C\circ b(i)=0$ for $i$ a letter, and that the
induction hypothesis \propGoal\ holds
for all $C\circ b(P)$ such that $\len{P}<\len{Q}$, so that
every application of the map $C$ can be removed from this equation. This leaves us with
\eqnn\propStepOneAndAHalf
$$\displaylines{
s_Q C\circ b(Q)=
\!\!\!\sum_{XY=Q}\!\!\!(k^X\cdot k^Y)\Big(b(X)\otimes b(Y)-b(Y)\otimes b(X)\Big)\hfil\propStepOneAndAHalf\hfilneg\cr
\sum_{XY=Q\atop \len{X}>1,\len{Y}>1}\!\!\!\!\sum_{AB=X}\!\!\!\big(b(A)\otimes b(B)-b(B)\otimes b(A)\big)\wedge b(Y)
+\sum_{XY=Q\atop \len{Y}=1}\sum_{CD=X}\!\!\!\big(b(C)\otimes b(D)-b(D)\otimes b(C)\big)\wedge b(Y)\cr
+\!\!\!\!\!\!\!\sum_{XY=Q\atop\len{X}>1,\len{Y}>1}\!\!\!\!\!\!\!b(X)\wedge\!\!\!\sum_{CD=Y}\!\!\!\big(b(C)\otimes b(D)-b(D)\otimes b(C)\big)
+\!\!\!\sum_{XY=Q\atop\len{X}=1}\!\!\!b(X)\wedge\!\!\sum_{AB=Y}\!\!\!\big(b(A)\otimes b(B)-b(B)\otimes b(A)\big)\cr
}$$
Absorbing the $\len{X}=1$ and $\len{Y}=1$ summations into the
$\len{X}>1$, $\len{Y}>1$ cases we get
\eqnn\propStepOneAndAHalfA
$$\displaylines{
s_Q C\circ b(Q)= \sum_{XY=Q}\!\!(k^X\cdot k^Y)\Big(b(X)\otimes b(Y)-b(Y)\otimes b(X)\Big)\hfil\propStepOneAndAHalfA\hfilneg\cr
+\!\!\! \sum_{XY=Q\atop \len{X}>1}\!\sum_{AB=X}\!\!\Big(b(A)\otimes b(B)-b(B)\otimes b(A)\Big)\wedge b(Y)
+\!\!\sum_{XY=Q\atop \len{Y}>1}\sum_{CD=Y}\!\!b(X)\wedge\Big(b(C)\otimes b(D)-b(D)\otimes b(C)\Big)\cr
}$$
Now we shall consider the two double sums. First of all we merge them
using that, for example, $\sum_{XY=Q,\len{X}>1}\sum_{AB=X}$ is the same as $\sum_{ABY=Q}$.
Then we remove the $\wedge$ using the definition \otdef\ to get
$$\displaylines{
\sum_{ABY=Q}\Big(b(A)\otimes b(B)-b(B)\otimes b(A)\Big)\wedge b(Y)
+\!\!\!\sum_{XCD=Q}b(X)\wedge\Big(b(C)\otimes b(D)-b(D)\otimes b(C)\Big)\cr
=\!\!\!\sum_{ABY=Q}\!\!\!\!\big([b(A),b(Y)]\otimes
b(B)+b(A)\otimes[b(B),b(Y)]-[b(B),b(Y)]\otimes b(A)-b(B)\otimes[b(A),b(Y)]\big)\cr
+\!\!\!\sum_{XCD=Q}\!\!\!\!\big([b(X),b(C)]\otimes b(D)+b(C)\otimes
[b(X),b(D)]-[b(X),b(D)]\otimes b(C)-b(D)\otimes [b(X),b(C)]\big)
}$$
We can now group the terms into two sets of four in a convenient way
\eqnn\propStepTwo
$$\eqalignno{
&= \Bigg(\sum_{ABY=Q}\Big([b(A),b(Y)]\otimes b(B)-b(B)\otimes[b(A),b(Y)]\Big)&\propStepTwo\cr
&\qquad{}\qquad{}\qquad{}\qquad{}\qquad{} +\sum_{XCD=Q}\Big(b(C)\otimes [b(X),b(D)]-[b(X),b(D)]\otimes b(C)\Big)\Bigg)\cr
&+\Bigg(\sum_{ABY=Q}\Big(b(A)\otimes[b(B),b(Y)]-[b(B),b(Y)]\otimes b(A)\Big)\cr
&\qquad{}\qquad{}\qquad{}\qquad{}\qquad{} +\sum_{XCD=Q}\Big([b(X),b(C)]\otimes b(D)-b(D)\otimes [b(X),b(C)]\Big)\Bigg)\cr
}$$
which we will now look at separately.
With the first set of terms, it is clear from relabeling the
second sum that it is just
$$
\sum_{ABY=Q}\!\!\!\Big([b(A),b(Y)]\otimes b(B)-b(B)\otimes[b(A),b(Y)] +b(B)\otimes [b(A),b(Y)]-[b(A),b(Y)]\otimes b(B)\Big)
$$
which is identically zero. The second set of terms in \propStepTwo\ can be
simplified using the definition of the $b$ map \BGbr\ leading to
\eqnn\propStepFour
$$\eqalignno{
&\sum_{ABY=Q}\!\!\Big(b(A)\otimes b(BY)s_{BY}-s_{BY}b(BY)\otimes b(A)\Big)&\propStepFour\cr
&\qquad{}\qquad{}\qquad{}\qquad{}\qquad{} +\sum_{XCD=Q}\!\!\Big(s_{XC}b(XC)\otimes b(D)-b(D)\otimes b(XC)s_{XC}\Big).
}$$
Then, since $B$ and $Y$ are adjacent everywhere they appear in the first sum, we can
condense them into a single word, and likewise for $X$ and $C$ in the second sum. This
leaves us with\foot{There should be a $\len{Y}>1$ in the first sum and a
$\len{X}>1$ in the second, as these words come from combining two words of non-zero length.
This can be left implicit since $s_P=0$ if $\len{P}=1$.}
\eqn\propStepFourA{
\sum_{XY=Q}\!\!s_Y\Big(b(X)\otimes b(Y) -b(Y)\otimes b(X)\Big)
+ \sum_{XY=Q}\!\!s_X\Big(b(X)\otimes b(Y)-b(Y)\otimes b(X)\Big)\,.
}
We now return to \propStepOne\ and, using that the double sum terms
are given by \propStepFourA, we finally obtain
\eqnn\propStepFive
$$\eqalignno{
C\circ b(Q) &=\frac{1}{s_Q}\sum_{XY=Q}\Big[(s_X+s_Y+(k^X\cdot
k^Y))\Big(b(X)\otimes b(Y)-b(Y)\otimes b(X)\Big)\Big]\cr
&=\sum_{XY=Q}\Big(b(X)\otimes b(Y)-b(Y)\otimes b(X)\Big)&\propStepFive
} $$
since $s_X+s_Y+(k^X\cdot k^Y)=s_{XY}$. Hence the result is proved. \qed

\proclaim Lemma 1. If $P$ has the form a left-to-right Dynkin bracket
$P=[[...[p_1,p_2],p_3],...],p_{\len{P}}]$,
\eqn\CofSpecialTopology{
C\circ P=\!\!\!\!\sum_{XjY=P\atop \delta (Y)=R\otimes S}\!\!\!\!(k^X\cdot k^j)\big[XR\otimes
jS-(X\leftrightarrow j)\big]\,,
}
where the deshuffle map $\d(Y)$ is defined in \deshuffle.
\par
\noindent{\it Proof.\/}
We use induction. From \recurPQ\
it follows
that $C\circ [1,2]=(k^1\cdot k^2)(1\otimes 2-2\otimes 1)$. We then suppose that
the relation \CofSpecialTopology\ is satisfied for the bracket $P$, and consider $C\circ [P,q]$,
where $q$ is a single letter.
\eqnn\CofSpecialTopInduction
$$\eqalignno{
C\circ [P,q]&=(C\circ P)\wedge q + P\wedge (C\circ q) + (k^P\cdot
k^q)(P\otimes q - q\otimes P)&\CofSpecialTopInduction\cr
&=\!\!\! \sum_{XjY=P\atop \delta (Y)=R\otimes S}\!\!\!(k^X\cdot k^j)\big(XR\otimes
jS-(X\leftrightarrow j)\big)\wedge q+(k^P\cdot k^q)(P\otimes q - q\otimes P)\cr
&=\!\!\! \sum_{XjY=P\atop \delta (Y)=R\otimes S}\!\!\!(k^X\cdot k^j)\big(XRq\otimes
jS+XR\otimes jSq-(X\leftrightarrow j)\big)
+(k^P\cdot k^q)(P\otimes q - q\otimes P)\cr
&=\!\!\!\sum_{XjY=P\atop \delta (Yq)=R\otimes S}\!\!\!(k^X\cdot k^j)\big(XR\otimes
jS-(X\leftrightarrow j)\big) +(k^P\cdot k^q)(P\otimes q - q\otimes P)\cr
&=\!\!\! \sum_{XjY=Pq\atop \delta (Y)=R\otimes S}\!\!\!(k^X\cdot k^j)\big(XR\otimes
jS-(X\leftrightarrow j)\big)
}$$
where $\delta$ is the deshuffle map \deshuffle.
Hence if \CofSpecialTopology\ is true for the Dynkin bracket $P$, it is true for the Dynkin bracket $[P,q]$,
and so by induction the result is proved. \qed

This result is important, as it shows that the general
redefinition formulae of this paper reduce to those
previously found in \Gauge\ when the multiplicity is less than six.

\newsubsubsec\newmapsec Contact term-like algorithms for simplifying redefinition terms

In this subsection a further pair of algorithms based around that of
contact terms \recurPQ\ will be defined, which will be useful when simplifying
the redefinition terms \HhatDef\ in the next section. The first of these will be
denoted $\tilde C$, and is defined by
\eqn\altRecurPQ{
\tilde C\circ i \equiv 0,\qquad
\tilde C\circ [A,B] \equiv \big(C\circ A\big)\twedge B
+ A\twedge \big(C\circ B\big)\,,
}
(note the $C$ map \recurPQ\ on the right-hand side) where $\twedge$ is defined by
\eqn\tildeotdef{
(A\otimes B)\twedge C \equiv [A,C]\otimes B\,,\qquad
A\twedge (B\otimes C) \equiv [A,B]\otimes C\,.
}
In addition we define a related algorithm $\tilde C'$ in terms of $\tilde C$,
\eqn\tildeCp{
\tilde C'\circ i \equiv 0,\quad
\tilde C'\circ [A,B]\equiv\tilde C \circ [A,B]-\half (k^A\cdot k^B)(A\otimes B - B\otimes A)\,.
}
The following notation, similar to that of \CVTdef, will be used with these maps
\eqn\CtildeAndCtildeprimeDef{
\tilde C\dlb K,S\drb\circ [P,Q] \equiv \dlb K,S\drb\circ\big(\tilde C\circ
[P,Q]\big)\,,\quad
\tilde C'\dlb K,S\drb\circ [P,Q] \equiv \dlb K,S\drb\circ\big(\tilde C'\circ [P,Q]\big)
}
where the double bracket $\dlb \cdot,\cdot \drb$ is defined in \wordRep.

\proclaim Lemma 2. The map $\tilde C$ satisfies
\eqn\CtildeIdentity{
\tilde C \circ [P,Q]=\!\!\!\!\!\sum_{XjY=P\atop \delta (Y)=R\otimes S}
\!\!\!\!\!(k^X\cdot k^j)\big([XR,Q]\otimes jS -(X\leftrightarrow j)\big) -
(P\leftrightarrow Q)\,,
}
for any Dynkin brackets $P$ and $Q$.
\par
\noindent{\it Proof.} To see this we use the identity \CofSpecialTopology\ as
follows,
\eqnn\CtildeIdentityProof
$$\displaylines{
\tilde C\circ [P,Q] =(C\circ P)\twedge Q + P\twedge (C\circ Q)\hfil\CtildeIdentityProof\hfilneg\cr
=\!\!\!\! \sum_{XjY=P\atop \delta (Y)=R\otimes S}\!\!\!\!\!(k^X\cdot k^j)\big(XR\otimes jS-(X\leftrightarrow j)\big)\twedge Q
+ P\twedge\!\!\!\!\sum_{XjY=Q\atop\delta (Y)=R\otimes S}\!\!\!\!\!(k^X\cdot k^j)\big(XR\otimes jS-(X\leftrightarrow j)\big)\cr
=\!\!\!\! \sum_{XjY=P\atop \delta (Y)=R\otimes S}\!\!\!\!\!(k^X\cdot k^j)\big([XR,Q]\otimes jS-(X\leftrightarrow j)\big)
+\!\!\!\!\sum_{XjY=Q\atop \delta (Y)=R\otimes S}\!\!\!\!\!(k^X\cdot k^j)\big([P,XR]\otimes jS-(X\leftrightarrow j)\big),
}$$
the second equality coming from the definition \tildeotdef. The result follows
after using the antisymmetry $[P,XR]=-[XR,P]$ in the final line. \qed

For illustrative examples of the $\tilde C$ map, see the appendix~\Ctildesec.

\newnewsec\ExpConstr Redefinitions of local multiparticle superfields

In this section we write down the redefinition algorithms to obtain
multiparticle superfields in the so-called BCJ gauge starting from both the Lorenz and
hybrid gauges with the most general bracketing configurations. The
characterization of these redefinitions as a gauge transformation was identified
in \Gauge\ and it will be reviewed and expanded in the next section.

\newsubsec\multiSYMsec Multiparticle superfields

It was shown in \refs{\EOMbbs,\Gauge} that the single-particle description admits a
generalization in terms of {\it multiparticle} superfields
$A^P_\a(x,\t)$, $A^P_m(x,\t)$, $W_P^\a(x,\t)$
and $F^P_{mn}(x,\t)$, which, for convenience, are collected in
the set $K_P$
\eqn\KPdef{
K_P \in \{A_\a^P(x,\t),\; A^m_P(x,\t),\; W^\a_P(x,\t),\; F^{mn}_P(x,\t)\}\,.
}
We will review two different ways to construct them below. At the same time we
will seamlessly fill some gaps in the discussions of \refs{\EOMbbs,\Gauge} by
utilizing the framework developed in the previous section.

\newsubsubsec\SYMgauges Multiparticle superfield in the Lorenz gauge

The generalization of the single-particle linearized superfields of
\RankOneEOM\ to an arbitrary number of labels follows from the local version
of the recursive solution to the non-linear wave equations \nonwave\ and can
be summarized by the following definition\foot{The Lorenz gauge discussion in
\Gauge\ is missing the definition of the general field-strength
$\Fhat^{mn}_{[P,Q]}$ while the definition of $\What_{[P,Q]}^\a$ is misleading
as $\lie_3\circ\What^\a_{[12,3]}\neq 0$ if one does not use momentum
conservation.}:

\proclaim Definition 6 (Lorenz gauge). Multiparticle super-Yang--Mills superfields in the
Lorenz gauge are defined starting with the multiplicity-one
superfields $\hat A_\a^i$, $\hat A_m^i$, $\What^\a_i$ and $\Fhat_i^{mn}$ and
recursively for arbitrary nested bracketings via
\eqnn\Lorenzdef
$$\eqalignno{
\hat{A}_\alpha^{[P,Q]}&=
-\half\big[\hat A_\a^P(k^P\cdot \hat A^Q) + \hat A^P_m(\gamma^m \hat
W^Q)_\a-(P\leftrightarrow Q)\big]
&\Lorenzdef \cr
\hat{A}_m^{[P,Q]}&=-\half\big[\hat A^P_m(k^P\cdot \hat A^Q) + 
\hat A^P_n \hat F_{mn}^Q -
(\hat W^P\gamma_m \hat W^Q)-(P\leftrightarrow Q)\big]\cr
\hat{W}^\alpha_{[P,Q]}&=
{1\over4}\hat F^P_{rs}(\gamma^{rs}\hat W^Q)^\a
-\half(k^P\cdot \hat A^Q)\hat W^\alpha_P
- \half \What^{m\a}_P A^m_Q
-(P\leftrightarrow Q)\cr
\hat F^{[P,Q]}_{mn} &=
- {1\over 2} \big[ \Fhat^{mn}_P (k_P \cdot \Ahat_Q)
+ \Fhat_{P}^{p|mn} \Ahat_p^Q + 2 \Fhat_P^{mp} \Fhat_{Q \, p}^n
+ 4 \g^{[m}_{\a\b} \What_P^{n]\a} \What_Q^\b
- (P \leftrightarrow Q) \big]\cr
}$$
where
\eqnn\Wmal
$$\eqalignno{
\hat{W}^{m\alpha}_{[P,Q]}&= k^m_{PQ}\What^\a_{[P,Q]} -
C\dlb\Ahat^m,\What^\a\drb\circ [P,Q]&\Wmal\cr
\Fhat^{m|pq}_{[P,Q]}&=k^m_{PQ}\Fhat^{pq}_{[P,Q]} -
C\dlb\Ahat^m,\Fhat^{pq}\drb\circ[P,Q]\,,\cr
}$$
and the map $C\circ$ is defined in \recurPQ.
Alternatively, the field-strength can be written as
\eqn\fsformFPQ{
\Fhat^{mn}_{[P,Q]} = k^m_{PQ}\Ahat^n_{[P,Q]}
- k^m_{PQ}\Ahat^m_{[P,Q]}
- C\dlb \Ahat^m,\Ahat^n\drb \circ[P,Q]\,.
}
\par
\noindent These recursions apply to arbitrary
bracketing structures encompassed by $P$ and $Q$. For example
$\Ahat^m_{[[1,2],[[3,4],5] ]}$
implies that $P=[1,2]$ and $Q=[[3,4],5]$ and leads to
\eqnn\exbra
$$\eqalignno{
\Ahat^m_{[[1,2],[[3,4],5] ]} =
-\half\Big[\hat A^{[1,2]}_m(k^{12}\cdot \hat A^{[[3,4],5]})& +
\hat A^{[1,2]}_n \hat F_{mn}^{[[3,4],5]}&\exbra\cr
& -(\hat W^{[1,2]}\gamma_m \hat W^{[[3,4],5]})-([1,2]\leftrightarrow
[[3,4],5])\Big]\,.
}$$
In addition, from the example for $C\circ[[1,2],[3,4]]$ in \Cexamples\
we have for \fsformFPQ,
\eqnn\Ftwotwo
$$\eqalignno{
\Fhat_{[[1,2],[3,4]]}^{mn} &=
k^m_{1234}\Ahat^n_{[[1,2],[3,4]]}
- k^n_{1234}\Ahat^m_{[[1,2],[3,4]]}&\Ftwotwo\cr
&{}-(k^1\cdot k^2)\big(
\Ahat_{[1,[3,4]]}^m \Ahat^n_2 +
\Ahat^m_1\Ahat^n_{[2,[3,4]]}
- (1\leftrightarrow2)\big) \cr
&{}-(k^3\cdot k^4)\big(
\Ahat^m_{[[1,2],3]}\Ahat^n_4
+ \Ahat^m_3\Ahat_{[[1,2],4]}
- (3\leftrightarrow4)\big) \cr
&{}-(k^{12}\cdot k^{34})\big( \Ahat^m_{[1,2]}\Ahat^n_{[3,4]} -
\Ahat^m_{[3,4]}\Ahat^n_{[1,2]}\big)\,.
}$$
Identifying the pair of words $P$ and $Q$ for the superfields on the right-hand side
of \exbra\ leads to further applications of the recursions in \Lorenzdef\ until
eventually all superfields are of single-particle nature.

\newsubsubsec\SYMBCJhybrid Multiparticle superfields in the hybrid gauge

Let us assume that all superfields of multiplicities $P$ and $Q$ in $K_P$ and
$K_Q$ have been redefined to satisfy all the BCJ symmetries \BCJcond\ (we will
explain how to do this below). Since multiparticle superfields $K_P$ in the BCJ
gauge satisfy the same symmetries as the Dynkin bracket $P=[[...[p_1,p_2],p_3],...],p_{\len{P}}]$
their multiparticle labels will be written as plain words $P=p_1p_2 \ldots
p_\len{P}$.
One then defines higher-multiplicity
superfields in $\check K_{[P,Q]}$ as follows:

\proclaim Definition 7 (Hybrid gauge). Multiparticle super-Yang--Mills superfields in the
hybrid gauge are distinguished by a check accent $\check K_{[P,Q]}$ and are defined by
\eqnn\Hybriddef
$$\eqalignno{
\check A_\alpha^{[P,Q]}&=
-\half[ A_\a^P(k^P\cdot  A^Q) +  A^P_m(\gamma^m  W^Q)_\a-(P\leftrightarrow Q)]
&\Hybriddef \cr
\check A_m^{[P,Q]}&=-\half[ A^P_m(k^P\cdot  A^Q) + 
 A^P_n  F_{mn}^Q -
( W^P\gamma_m  W^Q)-(P\leftrightarrow Q)]\cr
\check W^\alpha_{[P,Q]}&=
{1\over4} F^P_{rs}(\gamma^{rs} W^Q)^\a
-\half(k^P\cdot  A^Q) W^\alpha_P
- \half W^{m\a}_P A^m_Q
-(P\leftrightarrow Q)\cr
\check F^{[P,Q]}_{mn} &=
- {1\over 2} \big[ F^{mn}_P (k_P \cdot A_Q)
+ F_{P}^{p|mn} A_p^Q + 2 F_P^{mp} F_{Q \, p}^n
+ 4 \g^{[m}_{\a\b} W_P^{n]\a} W_Q^\b
- (P \leftrightarrow Q) \big]
}$$
where the superfields in $K_P$ and $K_Q$ on the right-hand side satisfy
the generalized Jacobi identities \BCJcond\ and
\eqnn\WmalUnhatted
$$\eqalignno{
{W}^{m\alpha}_{[P,Q]}&= k^m_{PQ}W^\a_{[P,Q]} -
C\dlb A^m,W^\a\drb\circ [P,Q]&\WmalUnhatted\cr
F^{m|pq}_{[P,Q]}&=k^m_{PQ}F^{pq}_{[P,Q]} -
C\dlb A^m,F^{pq}\drb\circ[P,Q]\,,\cr
}$$
are the local form of the superfields of higher-mass dimension defined in
\Gauge\ with the map
$C\dlb\cdot,\cdot\drb$ as in \CVTdef.
\par
\noindent Note an important difference with respect to the
definitions of superfields $\hat K_{[P,Q]}$ in the Lorenz gauge \Lorenzdef.
The definitions in the Lorenz gauge are recursive while
in the hybrid gauge they are {\it not} --
the superfields $\check K_{[P,Q]}$ on the left-hand side of
\Hybriddef\ have to be redefined before they can be used as the input on the
right-hand side at the next step. However, from a purely practical perspective,
to obtain the explicit expressions of the superfields in the BCJ gauge it is
more convenient to use the hybrid gauge.

\newsubsec\hybridsec  From hybrid gauge to BCJ gauge

The general formula to redefine the superfields
$\check K_{[P,Q]}\in \{ \check A_\alpha, \check A^m,\check W^\alpha \} $ from the hybrid gauge
\Hybriddef\ to superfields $K_{[P,Q]}\in \{ A_\alpha , A^m, W^\alpha \}$ in the BCJ gauge
is given by
\eqnn\Hpdef
$$\eqalignno{
K_{[P,Q]} &\equiv \check K_{[P,Q]} -
\!\!\!\!\sum_{P=XjY\atop \d(Y)=R\otimes S}\!\!\!\!\! (k_{X}\cdot k_j)
\bigl[ H_{[XR,Q]}\; K_{jS} - (X\leftrightarrow j)\bigr] &\Hpdef \cr
&+\!\!\!\! \sum_{Q=XjY\atop \d(Y)=R\otimes S}\!\!\!\!\! (k_{X}\cdot k_j)
\bigl[ H_{[XR,P]}\; K_{jS} - (X\leftrightarrow j)\bigr]
- \cases{ D_\alpha H_{[P,Q]} &: \ $K= A_\alpha$\cr
k_{PQ}^m H_{[P,Q]} &: \ $K= A^m$ \cr
0 &: \ $K= W^\alpha$}\,.
}$$
Alternatively, the identity \CtildeIdentity\ can be used to
rewrite \Hpdef\ more succinctly as\foot{It should be noted that,
despite \HpdefAlternative\ being defined for general
bracketing structures, it has only been verified for
$P$ and $Q$ Dynkin brackets in accordance with \Hpdef
.}
\eqn\HpdefAlternative{
K^{[P,Q]}=\check K^{[P,Q]}-\tilde C\dlb H, K\drb \circ [P,Q]-\cases{ D_\alpha H_{[P,Q]} &: \ $K= A_\alpha$\cr
k_{PQ}^m H_{[P,Q]} &: \ $K= A^m$ \cr
0 &: \ $K= W^\alpha$}\,.
}
These redefinitions introduce new superfields $H_{[P,Q]}$
whose purpose is to make the resulting linear combinations satisfy the BCJ
symmetries.
For example, the first instances of
the redefinition \Hpdef\ for $A^m_{[P,Q]}$ up to multiplicity
$\len{P}+\len{Q}=5$ are given by (recall that $\check A^m_i\equiv A^m_i$ and
$\check A^m_{[i,j]} \equiv A^m_{ij}$)
\eqnn\examples
$$\eqalignno{
A^m_{[1,2]} &=\check A^m_{[1,2]} &\examples\cr
A^m_{[12,3]} &= \check A^m_{[12,3]}-k_{123}^mH_{[12,3]}\cr
A^m_{[12,34]} &= \check A^m_{[12,34]} - (k^{1}\cdot k^{2})\Big[  H_{[1,34]}A^m_{2}-H_{[2,34]}A^m_1 \Big] \cr
&\quad{}+(k^{3}\cdot k^{4})\Big[  H_{[3,12]}A^m_{4}-H_{[4,12]}A^m_3 \Big] -k_{1234}^mH_{[12,34]}\cr
A^m_{[123,4]} &= \check A^m_{[123,4]} - (k^1\cdot k^2)\Big[ H_{[13,4]}A_2^m - H_{[23,4]}A_1^m\Big] \cr
&\quad{}-(k^{12}\cdot k^3)H_{[12,4]}A_3^m-k_{1234}^mH_{[123,4]}\cr
A^m_{[1234,5]} &= \check A^m_{[1234,5]}
- (k_1\cdot k_2) \big[
           H_{[134,5]} A^m_{2}
          + H_{[14,5]} A^m_{23}
          + H_{[13,5]} A^m_{24}
	  - (1\leftrightarrow 2)
          \big]\cr
\quad{}&       - (k_{12}\cdot k_3) \big[
          H_{[124,5]} A^m_{3}
          + H_{[12,5]} A^m_{34}
	  - (12\leftrightarrow 3)
          \big]\cr
\quad{}&       - (k_{123}\cdot k_4) H_{[123,5]} A^m_{4}
       - k^m_{12345} H_{[1234,5]}\cr
A^m_{[123,45]} &= \check A^m_{[123,45]} - (k^1\cdot k^2)\Big[ H_{[13,45]}A^m_{2} +
H_{[1,45]}A^m_{23} -(1\leftrightarrow 2)\Big] \cr
&\quad{}-(k^{12}\cdot k^3)\Big[ H_{[12,45]}A^m_3 -(12\leftrightarrow 3)\Big]\cr
&\quad{}+(k^4\cdot k^5)\Big[H_{[4,123]}A^m_5-(4\leftrightarrow 5)\Big] -k_{12345}^mH_{[123,45]}\cr
}$$

To help in elucidating the outcome of the above redefinitions we note that,
for suitable $H_{[P,Q]}$ to be given below, the superfields $K_{[P,Q]}$ on the left-hand side satisfy
all the identities implied by the bracket structure. For example,
\eqn\exSym{
A^m_{[12,3]} = - A^m_{[21,3]} = - A^m_{[3,12]}\,,\qquad
A^m_{[12,3]} + A^m_{[23,1]} + A^m_{[31,2]} = 0\,.
}
The above means that $A^m_{[[1,2],3]}$ satisfies the same symmetries as
$[[1,2],3]$ and can be represented via the shorthand $A^m_{123}\equiv A^m_{[[1,2],3]}$.
In general, the effect of the
above redefinitions is such that $K_{[P,Q]}=K_{P\ell(Q)}$, as shown in \Baker.

We have not yet discussed how the field strength $F_{[P,Q]}^{mn}$
superfields in the BCJ gauge are found.
These are most easily described by constructing them in
terms of the above redefined BCJ gauge superfields and using the contact-term
map \recurPQ,
\eqn\BCJgaugeFdef{
F_{[P,Q]}^{mn}=k^{PQ}_mA^{[P,Q]}_n-k^{PQ}_nA^{[P,Q]}_m-C\dlb A_m,A_n\drb \circ [P,Q]\,.
}

\newsubsubsec\HPQsols The explicit expression of $H_{[A,B]}$

In \Gauge\ the explicit form of the superfields $H_{[A,B]}$ was only given up
to multiplicity five. We now propose the following recursive solution for
general multiplicities\foot{We acknowledge the invaluable usage of {\tt FORM}
\FORM\ in these calculations.}
\eqn\genH{
H_{[i,j]}=0\,,\qquad H_{[A,B]}=
(-1)^{\len{B}}{\len{A}\over\len{A}+\len{B}}
\sum_{XjY=\dot a\tilde B}(-1)^{|Y|}H'_{\tilde Y,j,X} - (A\leftrightarrow B)\,,
}
where $\dot a$ and $\dot b$ denote the letterifications of $A$ and $B$ as defined in the
appendix~\Wordapp\ and
\eqnn\Hprimedef
\eqnn\HABCdef
$$\eqalignno{
H'_{A,B,C} &\equiv H_{A,B,C} +\Big[\half H_{[A,B]}(k_{AB}\cdot A_C)  + {\rm
cyclic}(A,B,C)\Big]&\Hprimedef\cr
& -\Big[\!\!\!\! \sum_{XjY=A\atop \delta(Y)=R\otimes S}
\!\!\!\!\!(k^{X}\cdot k^j)\big[H_{[XR,B]}H_{[jS,C]} -(X\leftrightarrow j)\big]
 + {\rm cyclic}(A,B,C)\Big]\,,\cr
H_{A,B,C} &\equiv -{1\over 4}A^m_A A^n_B F^{mn}_C
+ \half (W_A\g_m W_B)A_C^m + {\rm cyclic}(A,B,C)\,. &\HABCdef
}$$
Given that $H_{[A,B]}$ of multiplicities less than three vanish, it is easy to
see that the second line of \Hprimedef\ can only be probed when the
superfields have multiplicity six or higher. Furthermore, note that
$H_{[A,B]}$ satisfies generalized Jacobi identities within $A$ and $B$ and
therefore will be written using plain\foot{By convention, a plain word in a BCJ-gauge superfield
is a shorthand for the left-to-right nested
bracketing, e.g $P=1234\leftrightarrow P=[[[1,2],3],4]$.} words.

The superfields $H_{[P,Q]}$ up to multiplicity
seven are given by
\eqnn\explicitHs
$$\eqalignno{
H_{[12,3]}&=\frac{1}{3}\big(H'_{1,2,3}\big) &\explicitHs\cr
H_{[123,4]}&=\frac{1}{4}\big(H'_{12,3,4}-H'_{1,2,43}\big)\cr
H_{[12,34]}&=\frac{1}{4}\big(2H'_{1,2,34}-2H'_{3,4,12}\big)\cr
H_{[1234,5]}&=\frac{1}{5}\big(H'_{123,4,5}-H'_{12,3,54}+H'_{1,2,543}\big)\cr
H_{[123,45]}&=\frac{1}{5}\big(2H'_{12,3,45}-2H'_{1,2,453}-3H'_{4,5,123}\big)\cr
H_{[12345,6]}&=\frac{1}{6}\big(H'_{1234,5,6}-H'_{123,4,65}+H'_{12,3,654}-H'_{1,2,6543}\big)\cr
H_{[1234,56]}&=\frac{1}{6}\big(2H'_{123,4,56}-2H'_{12,3,564}+2H'_{1,2,5643}-4H'_{5,6,1234}\big)\cr
H_{[123,456]}&=\frac{1}{6}\big(3H'_{12,3,456}-3H'_{1,2,4563}-3H'_{45,6,123}+3H'_{4,5,1236} \big)\cr
H_{[123456,7]}&=\frac{1}{7} \big( H'_{12345,6,7}-H'_{1234,5,76}+H'_{123,4,765}-H'_{12,3,7654}+H'_{1,2,76543} \big)\cr
H_{[12345,67]}&=\frac{1}{7}\big(2H'_{1234,5,67}-2H'_{123,4,675}+2H'_{12,3,6754}-2H'_{1,2,67543}-5H'_{6,7,12345} \big)\cr
H_{[1234,567]}&=\frac{1}{7}\big( 3H'_{123,4,567}-3H'_{12,3,5674}+3H'_{1,2,56743}-4H'_{56,7,1234}+4H'_{5,6,12347}\big)\,,
}$$
while higher multiplicity examples can be easily generated using the general
formula \genH.
We have explicitly tested that the superfields up to and including
multiplicity nine following from the formulas \Hpdef\ and \genH\
satisfy the generalized Jacobi identities\foot{To simplify the algebra we tested
the bosonic components. Since the backbone of the recursion \genH\ is
given by the supersymmetric $H_{A,B,C}$ we believe that \genH\ also
leads to correct fermionic components.}. Since new
corrections cubic in $H_{[A,B]}$ could be present at multiplicity nine, the fact
that these formulas lead to superfields satisfying the BCJ symmetries
suggest that \genH\ is correct for arbitrary multiplicity.

\newsubsec\Lorenzsec From Lorenz gauge to BCJ gauge

Alternatively, one can generate superfields in the BCJ gauge by starting from
the superfields in
the Lorenz gauge obtained through the recursions \Lorenzdef. The redefinitions
are more involved in this case and one can show that
to obtain their BCJ gauge counterparts requires the
following iterated redefinition,
\eqn\genRedef{
K^{[P,Q]}=L_1\circ \hat K^{[P,Q]}\,,
}
where the operator $L_j$ is defined by
\eqn\Ajdef{
L_{j}\circ \hat K^{[P,Q]} \equiv \hat K^{[P,Q]}
- \frac{1}{j} C\dlb\hat H,L_{(j+1)}\circ\hat K\drb\circ[P,Q]
- \frac{1}{j}\cases{ D_\alpha \hat H_{[P,Q]} &: \ $K= A_\alpha$\cr
k_{PQ}^m \hat H_{[P,Q]} &: \ $K= A^m$ \cr
0 &: \ $K= W^\alpha$}\,,
}
while $C\circ\dlb\cdot,\cdot\drb$ is defined in \CVTdef. Notice that
$L_j\circ\hat K^{[P,Q]}$ gives rise to the action of the operator
$L_{(j+1)}\circ\hat K^{[A,B]}$ on the right-hand side with
$\len{A}+\len{B}<\len{P}+\len{Q}$. Therefore this is a iteration over the
index $j$ which eventually stops. As we will see below, the iteration built
into the redefinition \genRedef\ yields the infinite series of non-linear terms
present in the finite gauge transformation \toBCJ.

The examples \examples\ of redefinitions from the hybrid to BCJ gauge have the
following Lorenz to BCJ counterparts, using \genRedef\ and keeping all the
nested Lie brackets explicit
\eqnn\lorbcjex
$$\eqalignno{
A^m_{[1,2]} &= \Ahat^m_{[1,2]}\,, &\lorbcjex\cr
A^m_{[[1,2],3]} &= \Ahat^m_{[[1,2],3]} - k^m_{123}\hat H_{[[1,2],3]}\,,\cr
A^m_{[[1,2],[3,4]]} &=
  \Ahat^m_{[ [ 1 , 2 ] , [ 3 , 4 ] ]}
 -  (k_{1}\cdot k_{2})   \Big(
 \hat H_{[ 1 , [ 3 , 4 ] ]} \Ahat^m_{2} 
 -  \hat H_{[ 2 , [ 3 , 4 ] ]} \Ahat^m_{1}
\Big)\cr
& +  (k_{3}\cdot k_{4})   \Big(
 \hat H_{[ [ 1 , 2 ] , 4 ]} \Ahat^m_{3}
 -  \hat H_{[ [ 1 , 2 ] , 3 ]} \Ahat^m_{4}
\Big)
 - k^m_{1234}  \hat H_{[ [ 1 , 2 ] , [ 3 , 4 ] ]}\,,\cr
A^m_{[[[1,2],3],4]} &=
  \Ahat^m_{[ [ [ 1 , 2 ] , 3 ] , 4 ]}
-  (k_{1}\cdot k_{2})   \Big(
   \hat H_{[ [ 1 , 3 ] , 4 ]} \Ahat^m_{2} 
 -  \hat H_{[ [ 2 , 3 ] , 4 ]} \Ahat^m_{1} 
\Big)\cr
& -  (k_{12}\cdot k_{3})   \Big(
  \hat H_{[ [ 1 , 2 ] , 4 ]} \Ahat^m_{3}
\Big)
 -  (k_{123}\cdot k_{4})   \Big( 
  \hat H_{[ [ 1 , 2 ] , 3 ]} \Ahat^m_{4} 
\Big)
 - k^m_{1234} \hat H_{[ [ [ 1 , 2 ] , 3 ] , 4 ]}\,, \cr
A^m_{[[[[1,2],3],4],5]} &=
  \Ahat^m_{[ [ [ [ 1 , 2 ] , 3 ] , 4 ] , 5 ]}
 -  (k_{1}\cdot k_{2})   \Big(
   \hat H_{[ [ 1 , 3 ] , 4 ]} \Ahat^m_{[ 2 , 5 ]} 
 +  \hat H_{[ [ 1 , 3 ] , 5 ]} \Ahat^m_{[ 2 , 4 ]}
 +  \hat H_{[ [ 1 , 4 ] , 5 ]} \Ahat^m_{[ 2 , 3 ]}\cr
&\qquad{} +  \hat H_{[ [ [ 1 , 3 ] , 4 ] , 5 ]} \Ahat^m_{2}
- (1\leftrightarrow2)
\Big)\cr
& -  (k_{12}\cdot k_{3})   \Big(
   \hat H_{[ [ 1 , 2 ] , 4 ]} \Ahat^m_{[ 3 , 5 ]} 
 +  \hat H_{[ [ 1 , 2 ] , 5 ]} \Ahat^m_{[ 3 , 4 ]} 
 +  \hat H_{[ [ [ 1 , 2 ] , 4 ] , 5 ]} \Ahat^m_{3}
 - ([1,2]\leftrightarrow3)
\Big)\cr
& -  (k_{123}\cdot k_{4})   \Big(
   \hat H_{[ [ 1 , 2 ] , 3 ]} \Ahat^m_{[ 4 , 5 ]} 
 +  \hat H_{[ [ [ 1 , 2 ] , 3 ] , 5 ]} \Ahat^m_{4} 
\Big)\cr
& -  (k_{1234}\cdot k_{5})   \Big(
 \hat H_{[ [ [ 1 , 2 ] , 3 ] , 4 ]} \Ahat^m_{5}
\Big)
 -  \hat H_{[ [ [ [ 1 , 2 ] , 3 ] , 4 ] , 5 ]} k^m_{12345}\,,\cr
A^m_{[[[1,2],3],[4,5]]} &=
\Ahat^m_{[ [ [ 1 , 2 ] , 3 ] , [ 4 , 5 ] ]}
-  (k_{1}\cdot k_{2})   \Big(
\hat H_{[ 1 , [ 4 , 5 ] ]} \Ahat^m_{[ 2 , 3 ]}
 +  \hat H_{[ [ 1 , 3 ] , [ 4 , 5 ] ]} \Ahat^m_{2}
- (1\leftrightarrow2)
\Big)\cr
&-  (k_{12}\cdot k_{3})   \Big(
  \hat H_{[ [ 1 , 2 ] , [ 4 , 5 ] ]} \Ahat^m_{3}
 -  \hat H_{[ 3 , [ 4 , 5 ] ]} \Ahat^m_{[ 1 , 2 ]}
\Big)\cr
& -  (k_{123}\cdot k_{45})   \Big(
 \hat H_{[ [ 1 , 2 ] , 3 ]} \Ahat^m_{[ 4 , 5 ]}
\Big) \cr
& +  (k_{4}\cdot k_{5})   \Big(
 \hat H_{[ [ [ 1 , 2 ] , 3 ] , 5 ]} \Ahat^m_{4}
 -  \hat H_{[ [ [ 1 , 2 ] , 3 ] , 4 ]} \Ahat^m_{5}
\Big)- k^m_{12345}\hat H_{[ [ [ 1 , 2 ] , 3 ] , [ 4 , 5 ] ]}\,.
}$$
To illustrate \genRedef\ when there is more than one iteration, consider the
redefinition of the
superfield $\Ahat_m^{[[12,34],56]}$ to the BCJ gauge. It starts as
\eqnn\exPone
$$\eqalignno{
A_m^{[[12,34],56]} &= L_{1}\circ\Ahat_m^{[[12,34],56]} &\exPone\cr
& = \hat A_m^{[[12,34],56]}-k_m^{123456}\hat H_{[[12,34],56]}
- C\dlb\hat H,L_{2}\circ\Ahat^m\drb\circ [[12,34],56]
}$$
Using the definition of the $C\circ$ map from \recurPQ\ leads to
\eqnn\genRedefExampleOne
$$\eqalignno{
A_m^{[[12,34],56]}&=\hat A_m^{[[12,34],56]}-k_m^{123456}\hat H_{[[12,34],56]}& \genRedefExampleOne\cr
&- (k^1\cdot k^2)\Big(\big(L_2\circ\hat A_m^2\big)\hat H_{[[1,34],56]}
+\big(L_2\circ\hat A_m^{[2,34]}\big)\hat H_{[1,56]}\cr
&\hskip50pt +\big(L_2\circ\hat A_m^{[2,56]}\big)\hat H_{[1,34]}-(1\leftrightarrow 2)\Big)\cr
&- (k^{12}\cdot k^{34})\Big(\big(L_2\circ\hat A^{34}_m\big)\hat H_{[12,56]}-(12\leftrightarrow 34)\Big)\cr
&- (k^{1234}\cdot k^{56})\big(L_2\circ\hat A^{56}_m\big)\hat H_{[12,34]}\cr
&- (k^3\cdot k^4)\Big(\big(L_2\circ\hat A_m^4\big)\hat H_{[123,56]}
+\big(L_2\circ\Ahat_m^{[12,4]}\big)\hat H_{[3,56]}\cr
&\hskip50pt +\big(L_2\circ\hat A_m^{[4,56]}\big)\hat H_{[12,3]}-(3\leftrightarrow 4)\Big)\cr
&- (k^5\cdot k^6)\Big(\big(L_2\circ\hat A^6_m\big)\hat H_{[[12,34],5]}-(5\leftrightarrow 6)\Big)\,.
}$$
Note that on most of the terms the iteration stops since
$L_2\circ\hat A^i_m=\hat A^i_m$ and $L_2\circ\hat A^{ij}_m=\hat A^{ij}_m$.
The only remaining non-trivial action $L_2\circ\Ahat_m^P$ are on terms are of multiplicity three.
From \genRedef\ we obtain,
\eqn\exPtwo{
L_2\circ\hat A_m^{[12,3]}=\hat A_m^{[12,3]}-\half k_{123}^m\hat H_{[12,3]}\,,
\qquad
L_2\circ\Ahat_m^{[1,23]}=\hat A_m^{[1,23]}-\half k_{123}^m\hat H_{[1,23]}.
}
Plugging all of this into \genRedefExampleOne\ yields
\eqnn\genRedefExampleOnePartTwo
$$\eqalignno{
A_m^{[[12,34],56]}&=\hat A_m^{[[12,34],56]}-k_m^{123456}\hat H_{[[12,34],56]}&\genRedefExampleOnePartTwo\cr
&- (k^1\cdot k^2)\Big(\hat A_m^2\hat H_{[[1,34],56]}+\hat A_m^{[2,34]}\hat H_{[1,56]}+\hat A_m^{[2,56]}\hat H_{[1,34]}\cr
&\qquad{}\qquad{}-\frac{1}{2}k_m^{234}\hat{H}_{[2,34]}\hat H_{[1,56]}
-\frac{1}{2}k_m^{256}\hat H_{[2,56]}\hat H_{[1,34]}-(1\leftrightarrow 2)\Big)\cr
&- (k^{12}\cdot k^{34})\Big(\hat A^{34}_m\hat H_{[12,56]}-(12\leftrightarrow 34)\Big)\cr
&- (k^{1234}\cdot k^{56})\hat A^{56}_m\hat H_{[12,34]}\cr
&- (k^3\cdot k^4)\Big(\hat A_m^4\hat H_{[123,56]}+\hat A_m^{[12,4]}\hat H_{[3,56]}+\hat A_m^{[4,56]}\hat H_{[12,3]}\cr
&\qquad{}\qquad{}-\frac{1}{2}k_m^{124}\hat{H}_{[12,4]}\hat H_{[3,56]}
-\frac{1}{2}k_m^{456}\hat H_{[4,56]}\hat H_{[12,3]}-(3\leftrightarrow 4)\Big)\cr
&- (k^5\cdot k^6)\Big(\hat A^6_m\hat H_{[[12,34],5]}-(5\leftrightarrow 6)\Big)\,.
}$$
Higher-rank examples can be similarly generated from the recursion \Ajdef.

\newsubsubsec\Hhatsec Explicit form of $\hat H_{[P,Q]}$ for the Lorenz to BCJ gauge redefinition

Each $\hat H_{[P,Q]}$ is defined by enforcing the BCJ symmetry on the corresponding
superfield $K_{[P,Q]}$. It has been found that up to multiplicity eight that these can be
simplified as
\eqnn\HhatDef
$$\eqalignno{
\hat H_{[A,B]}&=\hat H'_{[A,B]}-\frac{1}{2}\tilde C\dlb \hat H, \hat H\drb
\circ  [A,B]\,,&\HhatDef\cr
\hat H'_{[A,B]}&=H_{[A,B]}-\frac{1}{2}\Big[(\hat H'_Ak^m_A-\tilde C'\dlb\hat
H^m, \hat H\drb\circ A) A^{B}_m-(A\leftrightarrow B)\Big]\,,\cr
\hat H'_{i}&=\hat H'_{[i,j]}=0\,,
}$$
where the $H_{[A,B]}$ are defined as they were in \genH\ - \HABCdef, and
$\hat{H}_A^m\equiv k_A^m\hat H_A$. Furthermore, the maps $\tilde C$ and
$\tilde C'$ are the variants of the contact-term map $C$ defined in the
section~\newmapsec.

To demonstrate the meaning of these maps we will now provide examples.
First of all note that the $\tilde C$ and $\tilde C'$ maps in \HhatDef\ are
both associated with pairs of $\hat H$ superfields, each of which requires 
three indices, and so these terms will only be non-zero when
$\len{A}+\len{B}\geq 6$.
Thus at lower multiplicities these relations reduce to
equation (3.15) of \Gauge, as the $\tilde C$ and $\tilde C'$ terms only
start contributing at multiplicity $6+$.
An example of the relations in this case is as follows:
\eqnn\Htfiveex
$$\eqalignno{
\hat H_{[[[1,2],3],[4,5]]}&=\hat H'_{[[[1,2],3],[4,5]]} &\Htfiveex\cr
&=H_{[[[1,2],3],[4,5]]}-\frac{1}{2}k^m_{123}\hat H'_{[[1,2],3]}A^{[4,5]}_m\cr
&=H_{[[[1,2],3],[4,5]]}-\frac{1}{2}H_{[[1,2],3]}(k_{123}\cdot A^{[4,5]})\,.
}$$
We will now outline an example of \HhatDef\ for the multiplicity six
redefinition term $\hat H_{[[[[1,2],3],[4,5]],6]}$, which should
demonstrate the formulae more clearly.
\eqn\Htsixstart{
\hat H_{[[[[1,2],3],[4,5]],6]}= \hat H'_{[[[[1,2],3],[4,5]],6]} -
\frac{1}{2}\tilde C \dlb \hat H,\hat H\drb \circ [[[[1,2],3],[4,5]],6]\,.
}
The expansion of the $\tilde C$ term above is given as the example
\CtildeRankSixExample\ in appendix~\Ctildesec, and from it we see that
\eqnn\Htsixmid
$$\eqalignno{
\tilde C \dlb \hat H,\hat H\drb \circ [[[[1,2],3],[4,5]],6]&=(k^1\cdot
k^2)\big(\hat H_{[[1,3],6]}\hat H_{[2,[4,5]]} - \hat H_{[[2,3],6]}\hat H_{[1,[4,5]]}\big)&\Htsixmid\cr
&+(k^{12}\cdot k^3)\big( \hat H_{[[1,2],6]}\hat H_{[3,[4,5]]}\big) + (k^{123}\cdot k^{45})\big(\hat H_{[[4,5],6]}\hat H_{[[1,2],3]}\big)\cr
&=(k^1\cdot k^2)\big( H_{[[1,3],6]} H_{[2,[4,5]]} - H_{[[2,3],6]} H_{[1,[4,5]]}\big)\cr
&+(k^{12}\cdot k^3)\big(  H_{[[1,2],6]} H_{[3,[4,5]]}\big) + (k^{123}\cdot k^{45})\big( H_{[[4,5],6]} H_{[[1,2],3]}\big) 
}$$
As for the $\hat H'_{[[[[1,2],3],[4,5]],6]}$ term, this piece is given by
\eqnn\Htsixmidt
$$\eqalignno{
\hat H'_{[[[[1,2],3],[4,5]],6]}&=H_{[[[[1,2],3],[4,5]],6]}
-\frac{1}{2}\big[(\hat H'_{[[[1,2],3],[4,5]]}k_{12345}^m
-\tilde C'\dlb\hat H,\hat H\drb\circ [[[1,2],3],[4,5]])A^6_m\big]\cr
&=H_{[[[[1,2],3],[4,5]],6]}-\frac{1}{2}H_{[[[1,2],3],[4,5]]}(k_{12345}^m\cdot A^6)\cr
&\qquad{}+\frac{1}{4}H_{[[1,2],3]}(k_{123}\cdot A^{45})(k^{12345}\cdot A^6)\,, &\Htsixmidt
}$$
where we have used \Htfiveex\ and
that the action of $\tilde C'\dlb \hat H,\hat H\drb$ on
any Lie polynomial with less than six letters is zero. Putting this all
together we thus have that
\eqnn\Htsixend
$$\eqalignno{
\hat H_{[[[[1,2],3],[4,5]],6]}&=H_{[[[[1,2],3],[4,5]],6]}&\Htsixend\cr
&-\frac{1}{2}H_{[[[1,2],3],[4,5]]}(k_{12345}^m\cdot A^6)+\frac{1}{4}H_{[[1,2],3]}(k_{123}\cdot A^{45})(k^{12345}\cdot A^6) \cr
&-\frac{1}{2}(k_1\cdot k_2)\big( H_{[[1,3],6]} H_{[2,[4,5]]} - H_{[[2,3],6]} H_{[1,[4,5]]}\big)\cr
&-\frac{1}{2}(k_{12}\cdot k_3)\big(  H_{[[1,2],6]} H_{[3,[4,5]]}\big)
-\frac{1}{2} (k_{123}\cdot k_{45})\big( H_{[[4,5],6]} H_{[[1,2],3]}\big)\,.
}$$
Unfortunately to see an example where the $\tilde C'$ map in the definition
of $\hat H'$ comes into affect requires going to multiplicity seven, which considerably
increases the number of terms involved and makes any such example less easy to follow.
The process is not terribly different from the one just outlined though, there 
are just more terms involved.

It might raise some concerns that \HhatDef\ and \genH\ - \HABCdef\ are in some
places defined in terms of BCJ gauge superfields, and so this might not
represent a true gauge transformation. This is however not an issue, as a
purely Lorenz gauge version of \HhatDef\ can be found by just replacing the
BCJ superfields with their Lorenz gauge expansions \genRedef. Some difficulty
may arise doing this for $H_{A,B,C}$ due to the presence of $F^{mn}_P$ terms.
However, we do the same thing, and plug the Lorenz gauge expansions into
\BCJgaugeFdef\ to get
\eqn\BCJgaugeF{
F_{[P,Q]}^{mn}=
k^{PQ}_m(L_1\circ \hat A^{[P,Q]}_n)
-k^{PQ}_m(L_1\circ \hat A^{[P,Q]}_m)
-C\dlb (L_1\circ \hat A_m),(L_1\circ \hat A_n)\drb \circ [P,Q]\,.
}
The notation of \HhatDef\ has just been chosen for its compactness and clarity.

\newnewsec\BGHs BCJ symmetries and standard gauge transformations

In this section we will briefly review the result of \Gauge\ that the
redefinitions of a local superfield $K$ from the Lorenz to the BCJ gauge
amount to a standard gauge transformation of the corresponding non-linear
superfield $\Bbb K$ introduced in section~\nonlinSYMsec. However, the
discussion of \Gauge\ was based on examples up to multiplicity five and
consequently missed an infinite number of correction terms. As a result, the
gauge transformations were identified only in infinitesimal form. We will
prove that the iterative redefinitions \recurPQ\ lead to a {\it finite} gauge
transformation instead.

To show this one uses the perturbiner series expansion
$\Bbb K$ as given in \series\ in terms of its Berends--Giele currents.
Before proceeding, we review the definition of the Berends--Giele currents
using a formulation based on the $b$ map \BGmap.

\newsubsec\Cmaps Berends--Giele currents and contact terms from maps on words

We will define the notion of a {\it Berends--Giele current} from a purely
combinatorial point of view based on the map $b(P)$ acting on words.
In order to do this for arbitrary labelled objects such as multiparticle superfields,
let us define the a replacement of words by arbitrary
superfields as
\eqn\wordRep{
\dlb K\drb \circ P \equiv K_{P}\,,\qquad
\dlb K,S\drb \circ P\otimes Q \equiv K_P S_Q\,.
}
In turn, this definition can be used to define
the Berends--Giele currents and related concepts through
the $b$ and $C$ maps.

\proclaim Definition 4 (Berends--Giele map).
If $K_P\in\{A^P_\a,A^m_P,W^\a_P,F^{mn}_P\}$ is a local multiparticle superfield, its associated Berends--Giele current
is represented by a calligraphic letter $\cK_P\in
\{\cA^P_\a,\cA^m_P,\cW^\a_P,\cF^{mn}_P\}$ and is given by
\eqn\BGmap{
\cK_P \equiv \dlb K\drb\circ b(P)\,,
}
where $\dlb\cdot\drb$ is defined in \wordRep.
\par
\noindent For example, the Berends--Giele currents up to multiplicity five associated to the vector
potential $A^m_P$ following from the definition $\cA^m_P=\dlb A^m\drb\circ b(P)$
are given by $\cA^m_1 = A^m_1$ and
\eqnn\BGAms
$$\eqalignno{
\cA^m_{12} &= {A^m_{[1,2]}\over s_{12}}\,,&\BGAms\cr
\cA^m_{123} &= {A^m_{[ [ 1 , 2 ] , 3 ]} \over s_{12} s_{123}}
+  {A^m_{[ 1 , [ 2 , 3 ] ]} \over s_{123} s_{23}}\,,\cr
\cA^m_{1234} &=
{A^m_{[ [ [ 1 , 2 ] , 3 ] , 4 ]} \over s_{12} s_{123} s_{1234}}
+ {A^m_{[ [ 1 , [ 2 , 3 ] ] , 4 ]} \over s_{123} s_{1234} s_{23}}
+  {A^m_{[ [ 1 , 2 ] , [ 3 , 4 ] ]} \over s_{12} s_{1234} s_{34}}
+  {A^m_{[ 1 , [ [ 2 , 3 ] , 4 ] ]} \over s_{1234} s_{23} s_{234}}
+  {A^m_{[ 1 , [ 2 , [ 3 , 4 ] ] ]} \over s_{1234} s_{234} s_{34}}\,,\cr
\cA^m_{12345} &=
{A^m_{[ [ [ [ 1 , 2 ] , 3 ] , 4 ] , 5 ]} \over s_{12} s_{123} s_{1234} s_{12345}}
 +  {A^m_{[ [ [ 1 , [ 2 , 3 ] ] , 4 ] , 5 ]} \over s_{123} s_{1234} s_{12345} s_{23}}
 +  {A^m_{[ [ [ 1 , 2 ] , [ 3 , 4 ] ] , 5 ]} \over s_{12} s_{1234} s_{12345} s_{34}}
 +  {A^m_{[ [ [ 1 , 2 ] , 3 ] , [ 4 , 5 ] ]} \over s_{12} s_{123} s_{12345}  s_{45}}\cr
& +  {A^m_{[ [ 1 , [ [ 2 , 3 ] , 4 ] ] , 5 ]} \over s_{1234} s_{12345} s_{23} s_{234}}
 +  {A^m_{[ [ 1 , [ 2 , [ 3 , 4 ] ] ] , 5 ]} \over s_{1234} s_{12345} s_{234} s_{34}}
 +  {A^m_{[ [ 1 , [ 2 , 3 ] ] , [ 4 , 5 ] ]} \over s_{123} s_{12345} s_{23} s_{45}}
 +  {A^m_{[ [ 1 , 2 ] , [ [ 3 , 4 ] , 5 ] ]} \over s_{12} s_{12345} s_{34}
 s_{345}}\cr
& +  {A^m_{[ [ 1 , 2 ] , [ 3 , [ 4 , 5 ] ] ]} \over s_{12} s_{12345} s_{345} s_{45}}
 +  {A^m_{[ 1 , [ [ [ 2 , 3 ] , 4 ] , 5 ] ]} \over s_{12345} s_{23} s_{234} s_{2345}}
 +  {A^m_{[ 1 , [ [ 2 , [ 3 , 4 ] ] , 5 ] ]} \over s_{12345} s_{234} s_{2345} s_{34}}
 +  {A^m_{[ 1 , [ [ 2 , 3 ] , [ 4 , 5 ] ] ]} \over s_{12345} s_{23} s_{2345}
 s_{45}}\cr
& +  {A^m_{[ 1 , [ 2 , [ [ 3 , 4 ] , 5 ] ] ]} \over s_{12345} s_{2345} s_{34} s_{345}}
 +  {A^m_{[ 1 , [ 2 , [ 3 , [ 4 , 5 ] ] ] ]} \over s_{12345} s_{2345} s_{345} s_{45}}\,.
}$$
The multiplicity six case is given in equation \RankSixBGCurrent\ of the
appendix. Moreover, one can show that $M_P=\dlb V\drb\circ b(P)$ reproduces
the intuitive Berends--Giele definition given in the appendix of \EOMbbs.
See~\BGfour.

\newsubsec\BGgaugetransec BCJ symmetries of local superfields as a gauge transformation

It was already pointed out in \Gauge\ that the redefinitions
of the local multiparticle superfields in the Lorenz gauge correspond to a gauge
transformation of the corresponding Berends--Giele current.

Indeed, if we define the Berends--Giele currents using \BGmap
\eqn\BGAH{
\cA_P^m \equiv \dlb A^m\drb\circ b(P)\,,\qquad
\cH \equiv \dlb H\drb\circ b(P)\,,
}
one can show using the relations \lorbcjex\ and \BGAms\ up to multiplicity five that \Gauge,
\eqnn\bcjgauges
$$\eqalignno{
\cA^{m,\rm BCJ}_{123} &= \cA^{m,\rm L}_{123} - k^m_{123}\cH_{123}\,,&\bcjgauges\cr
\cA^{m,\rm BCJ}_{1234} &= \cA^{m,\rm L}_{1234} - k^m_{1234}\cH_{1234}+
\cA^{m,\rm L}_1\cH_{234} - \cA^{m,\rm L}_4\cH_{123}\,,\cr
\cA^{m,\rm BCJ}_{12345} &= \cA^{m,\rm L}_{12345} - k^m_{12345}\cH_{12345}
+ \cA^{m,\rm L}_1\cH_{2345}
+ \cA^{m,\rm L}_{12}\cH_{345}
- \cA^{m,\rm L}_5\cH_{1234}
- \cA^{m,\rm L}_{45}\cH_{123}\,.
}$$
Therefore, in terms of the perturbiner series
\eqn\Hseries{
\Bbb H \equiv \sum_P \cH_P t^P\,,
}
the equations \bcjgauges\ correspond to the infinitesimal non-linear gauge
transformation \NLgauge\ with $\Omega=-\Bbb H$
\eqn\infinitesimal{
\Bbb A_m^{\rm BCJ} = \Bbb A^{\rm L}_m - [\p_m,\Bbb H] + [\Bbb A^{\rm L}_m,\Bbb H]\,.
}
However, the identification of \infinitesimal\ as the gauge transformation
relating the superfields in the different gauges is not complete. This is
because the analysis of \Gauge\ was restricted to multiplicity five, whereas
we know from \genH\ and \Hprimedef\ that there are non-linear corrections to the
superfields $H_{[A,B]}$ that start at multiplicity six -- see for instance the
quadratic terms $\sim \half k^m H^2$ in the
redefinition of $\Ahat^{[[12,34],56]}_m$ \genRedefExampleOnePartTwo.

In fact, using the general formulas for the redefinitions and the
Berends--Giele currents one can show, after considerable effort, 
\eqnn\bcjsixgauge
$$\eqalignno{
\cA^{m,\rm BCJ}_m & =\cA^{m,\rm L}_{123456} - k^m_{123456}\cH_{123456}&\bcjsixgauge\cr
&+ \cA^{m,\rm L}_1\cH_{23456}
+ \cA^{m,\rm L}_{12}\cH_{3456}
+ \cA^{m,\rm L}_{123}\cH_{456}
- \cA^{m,\rm L}_6\cH_{12345}
- \cA^{m,\rm L}_{56}\cH_{1234}
- \cA^{m,\rm L}_{456}\cH_{123}\cr
&-\half k^m_{123}\cH_{123}\cH_{456}+ \half k^m_{456}\cH_{456}\cH_{123}\,.
}$$
Therefore, at multiplicity six the transformation between Lorenz and BCJ gauge
follows from
\eqn\infiplus{
\Bbb A_m^{\rm BCJ} = \Bbb A^{\rm L}_m - [\p_m,\Bbb H] + [\Bbb A^{\rm L}_m,\Bbb H]
- \half[[\p_m,\Bbb H],\Bbb H]\,.
}
We will now demonstrate that there is an infinite series of non-linear corrections to
\infiplus\ which generate a {\it finite} gauge variation.

\newsubsec\gaugetranssec BCJ symmetries from finite gauge transformations

If $\Bbb H$ represents a generating series of Berends--Giele
superfields $\cal H_P$ \Hseries, one can show that the
series representation of the recursive iterations \Ajdef\ for the gauge
superpotential $\Bbb A_m$ is given by
\eqn\Lseries{
{\Bbb L}_j\circ\Bbb A_m = \Bbb A_m - {1\over j}[\p_m,\Bbb H] - {1\over j}[\Bbb
H, {\Bbb L}_{j+1}\circ \Bbb A_m]\,.
}
Iterating the series representation of the transformation $\Bbb A_m^{\rm BCJ}
= \Bbb L_1\circ\Bbb A^{\rm L}_m$
from Lorenz to BCJ gauge leads to ($\nabla^{\rm L}_m \equiv \p_m - \Bbb A_m^{\rm L}$)
\eqnn\toBCJ
$$\eqalignno{
\Bbb A_m^{{\rm BCJ}} &=
\Bbb A_m^{{\rm L}}
+ [\Bbb H,\p_m] - [\Bbb H, \Bbb A_m^{{\rm L}}]
- \half [\Bbb H, [\Bbb H,\p_m]] + \half [\Bbb H,[\Bbb H, \Bbb A^{\rm L}_m]]
+ {1\over 3!} [\Bbb H,[\Bbb H, [\Bbb H,\p_m]]]
+ \cdots\cr
&=\Bbb A_m^{{\rm L}}
+ [\Bbb H,\nabla^{\rm L}_m]
- \half [\Bbb H, [\Bbb H,\nabla^{\rm L}_m]]
+ {1\over 3!} [\Bbb H,[\Bbb H, [\Bbb H,\nabla^{\rm L}_m]]] + \cdots &\toBCJ
}$$
Unsurprisingly, the expression \toBCJ\
is nothing more than the series expansion of the {\it finite}
gauge transformation given by
\eqn\finitegau{
\Bbb A_m^{\rm BCJ} = U\Bbb A_m^{\rm L} U^{-1} + \p_mU U^{-1}\,,\quad U =
\exp(-\Bbb H)\,.
}
Alternatively \toBCJ\ can be rewritten as
$\nabla_m^{\rm BCJ} = e^{-{\rm ad}_{\Bbb H}}(\nabla^{\rm L}_m)$,
where ${\rm ad}_{\Bbb H}(X) \equiv [\Bbb H,X]$.

\newnewsec\outlooksec Conclusions and outlook
\global\subsecno=1

One of the main achievements of this paper is the recursive solution to the
redefinition superfields $H_{[A,B]}$ given in \genH. These superfields encode
the non-linear gauge variations required to obtain local multiparticle
superfields in the BCJ gauge. The pursuit of this formula led to improvements
to and clarifications of earlier discussions given in \refs{\EOMbbs,\Gauge}.
In particular, in going beyond the multiplicity-five examples of \Gauge, we
found an infinite set of higher-order corrections leading to the perturbiner
representation of a {\it finite} gauge transformation \toBCJ.

We also introduced new combinatorial maps on words and rigorously proved key
statements that address some natural although not crucial questions previously
left unanswered. For instance, we found closed formulas for the gauge
redefinition of $K_{[P,Q]}$ for arbitrary nested bracketings as well as the
field-strength form of $F^{mn}_{[P,Q]}$ and related superfields at higher-mass
dimension. Several other formulas along these lines can now be written down,
such as the local equations of motion \localEOM\ for the Lorenz-gauge
superfields $\hat K_{[P,Q]}$, again for arbitrary Lie bracket structure. The
precise definition of maps in section~\contactapp\ ultimately related to the
definition of Berends--Giele currents also lead to explanations of {\it why}
some patterns are ubiquitous when discussing BRST variations of various
superfields in the pure spinor formalism as seen in the discussions of
\oneloopI.

We will end this paper with some observations that could lead to
further investigations.

\newsubsubsec\curisec Tree-level amplitudes using redefinition superfields

The gauge transformations responsible for the BCJ gauge require redefinitions
by superfields of ghost-number zero $H_{[A,B]}$ determined recursively through
\genH. Customarily, after performing the redefinitions using the redefining
superfields one  writes down the tree amplitudes of SYM using the newly
obtained superfields \BGPS.
For example, using the compact language of the pure spinor superspace \exPSS\ one gets
 \eqn\fiveV{
 \AYM(1,2,3,4,5) =
  {\langle V_{123} V_{4} V_{5}\rangle\over s_{12}s_{123}}
+ {\langle V_{321} V_{4} V_{5}\rangle\over s_{23}s_{123}}
+ {\langle V_{12} V_{34} V_{5}\rangle\over s_{12}s_{34}}
+ {\langle V_{1} V_{432} V_{5}\rangle\over s_{34}s_{234}}
+ {\langle V_{1} V_{234} V_{5}\rangle\over s_{23}s_{234}}\,,
}
where $V_P\equiv \l^\a A_\a^P$ is a BCJ-satisfying superfield whose explicit
expression contains the redefinition superfields $H'_{A,B,C}$ in various
combinations.

So, in the usual formulation, we see that the superfields in the BCJ gauge are
used to write down the local numerators of tree-level SYM amplitudes. These
numerators have ghost number three \psf\ and, if one wishes to produce expressions
written in terms of particle polarizations and momenta, require the standard
pure spinor zero-mode rule
$\langle(\l\g^m\t)(\l\g^n\t)(\l\g^p\t)(\t\g_{mnp}\t)\rangle=1$ \psf\ to
integrate out the pure spinors. Somewhat surprisingly, it turns out that the
redefinition superfields themselves give rise to
numerators of the tree amplitudes of SYM.

\newsubsubsec\AYMmapsec Tree-level amplitudes as a map on planar binary trees

The observation above can be made more intuitive and intriguing if we frame it in
terms of the $b$ map \BGbr. The SYM tree amplitudes can be viewed as a map
$\AYM\circ$
acting on the Lie polynomials in the expansion of \BGbr. More precisely,
\eqn\SYMbmap{
\AYM(P,n) = s_P \AYM\circ \big(b(P)b(n)\big),
}
where the map $\AYM\circ$ admits two formulations
\eqn\AYMmap{
\AYM\circ [P,Q]n \equiv
\cases{ \langle V_PV_QV_n\rangle &\cr
H'_{P,Q,n} &\cr
}}
For example, using the Lie bracket expansion from~\figBGfour\ and the
top line of the map \AYMmap\ gives rise to amplitude expression \fiveV.
Using the bottom line of the map yields instead
\eqnn\fiveex
$$\eqalignno{
\AYM(1,2,3,4,5) &= s_{1234}\AYM\circ b(1234)b(5) &\fiveex\cr
&=
{H'_{123,4,5}\over s_{12}s_{123}}
+ {H'_{321,4,5}\over s_{23}s_{123}}
+ {H'_{12,34,5}\over s_{12}s_{34}}
+ {H'_{1,432,5}\over s_{34}s_{234}}
+ {H'_{1,234,5}\over s_{23}s_{234}}
}$$
In hindsight, the statement that tree-level amplitudes
can be written using the definition of $H_{A,B,C}$ could be made when putting
together the results of \BGPS\ and \Gauge. But now we have explicitly
checked up to multiplicity nine that
all the new corrections introduced in \Hprimedef\ that lead to the definition
of $H'_{A,B,C}$ do not affect the final results of the amplitudes.

These observations give rise to the speculation that the new prescription to
compute tree level amplitudes from \Newnathan\ naturally gives rise to the
amplitudes written in terms of $H'_{A,B,C}$. After all the prescription in \Newnathan\ does
not involve unintegrated vertices (so no pure spinors) and the end result
will have to involve the double poles in the OPEs among integrated vertices.
This agrees with the mechanism in the usual formulation \EOMbbs\ 
where the double poles are distributed among the simple poles using
integration by parts, and it is after this step that the superfields in the
numerators satisfy BCJ symmetries.
This may give rise to a systematic derivation of the $H'_{A,B,C}$
redefinitions via OPE calculations and it is an interesting question left to the
future.

BCJ numerators were constructed for gauge theories deformed by $\ap F^3$ and
$\ap^2 F^4$ interactions by finding appropriate $\ap$ corrections to the $H_P$
fields \BGFthree. Since low-multiplicity examples show that these corrections can also be written in terms of
$\ap$-corrected $H_{A,B,C}$ in a similar manner as discussed in this paper,
one may wonder whether the all-multiplicity formulas found here can be applied
with minimal changes to the setup of \BGFthree.

The color-kinematics duality has given reasons to speculate about the
existence of a ``kinematic algebra'' \kinalMon\ in the same way as the color
factors are related to standard Lie algebras. It will be interesting to connect
this line of thought with the gauge variation approach pursued here. See
\kinalJ\ for a recent account on the quest for the kinematic algebra.

Finally, the Berends--Giele recursion relations have been recently derived using the
technology of an $L_\infty$-algebra in \linftyW. It would be interesting to
find a new derivation of the recursions for the
gauge parameter $H_{[A,B]}$ using the methods of \linftyW.

\bigskip
\noindent{\bf Acknowledgements:}
EB thanks Kostas Skenderis for useful discussions. 
CRM thanks Oliver Schlotterer for collaboration on closely related topics and
for comments on the draft. CRM is supported by a University Research
Fellowship from the Royal Society.

\appendix{A}{Some common operations on words}
\applab\Wordapp

\noindent In this appendix we list some of the operations on words used in
this paper. With the exception of the letterification introduced below, the
following definitions
are standard and can be found in \Reutenauer.

The left-to-right bracketing map $\ell (A)$ is defined recursively by
\eqn\lrbracket{
\ell(123..n)\equiv
\ell(123...n-1)n-n\ell(123...n-1),\;\;\;\ell(i)=i,\;\;\;\ell(\emptyset)=0\,.
}
The deshuffle map is defined by
\eqn\deshuffle{
\d(P) = \sum_{X,Y}\langle P, X\shuffle Y\rangle\, X\otimes Y\,,
}
where $\langle \cdot,\cdot\rangle$ denotes
the scalar product on words
\eqn\AdotB{
\langle A, B\rangle \equiv \d_{A,B},\qquad
\d_{A,B}= \cases{$1$, & if $A=B$\cr
		 $0$, & otherwise}\,.
}
The shuffle product $\shuffle$ between $A=a_1 a_2
\ldots a_{|A|}$ and $B= b_1 b_2\ldots b_{|B|}$ is given by
\eqn\shuffledef{
\emptyset\shuffle A = A\shuffle\emptyset = A,\qquad
A\shuffle B \equiv a_1(a_2 \ldots a_{|A|} \shuffle B) + b_1(b_2 \ldots b_{|B|}
\shuffle A)\,,
}
where $\emptyset$ represents the empty word.

In certain formulas such as \genH\ it is necessary to handle a word as if it were
a single letter to avoid it being split by other maps. To deal with these
situations we introduce a {\it letterfication} operation whereby a
{\it word\/} $Q$ is mapped to a {\it letter\/} $\dot q$,
\eqn\letterif{
Q\rightarrow \dot q\,.
}
Since a letter can not be
deconcatenated this freezes the individual letters within $Q$. In the end $\dot q$ is
restored by its original word $Q$. For example, suppose that the word $Q=12$
has been letterified to $\dot q = 12$ -- as may be the case in a formula such as
\genH -- and that $P=3$. Then deconcatenating $QP$ is different
than deconcatenating $\dot qP$. For example, one gets only one term
\eqn\exlett{
Q=12, P=3\rightarrow \sum_{XY=\dot qP}S_X T_Y = S_{\dot q}T_3 = S_{12}T_{3}
}
instead of the usual two ($S_1T_{23} + S_{12}T_3$) if $Q$ is not letterified.

\appendix{B}{Equations of motion for local $\hat K_{[P,Q]}$}
\applab\eomapp

\noindent In this appendix we will write down the equations of motion satisfied by the
multiparticle superfields in the Lorenz gauge for general nested Lie brackets.

The equations of motion satisfied by the local multiparticle superfields
\Lorenzdef\ can be written as a local counterpart of the non-linear equations \SYMeom
\eqn\localEOM{
\eqalign{
\nabla_{(\a}^{(L)}\Ahat_{\b)}^{[P,Q]} &= \g^m_{\a\b} \Ahat^m_{[P,Q]} \cr
\lnabla_\a \hat A^m_{[P,Q]} &=  (\g^m  \What_{[P,Q]})_\a + k^m_{PQ}  \Ahat_{\a\, [P,Q]}
}\qquad\eqalign{
\lnabla_\a \What^\b_{[P,Q]} &=  {1\over 4}(\g^{mn})_\a{}^\b \Fhat^{[P,Q]}_{mn}\cr
\lnabla_\a \Fhat^{mn}_{[P,Q]} &= \big(\What_{[P,Q]}^{[m}\g^{n]}\big)_\a
}}
where $\lnabla_\a$ is the local counterpart of $\nabla_\a \equiv D_\a - \Bbb
A_\a$ and is defined by
\eqn\localnabladef{
\lnabla_\a \equiv D_\a - C\dlb\Ahat_\a,\,\cdot\;\drb\,,\qquad
C\dlb\Ahat_\a,\cdot\;\drb K_{[P,Q]}\equiv C\dlb\Ahat_\a,K\drb\circ[P,Q]\,.
}
where $C\dlb\cdot,\cdot\drb$ is the contact-term coproduct map on words
defined in \recurPQ\ and \CVTdef.
To illustrate the above equations, consider
$\lnabla_\a \Ahat^m_{[1,2]} = (\g^m  \What_{[1,2]})_\a + k^m_{12}  \Ahat_{\a\,[1,2]}$
where
\eqnn\Firstex
$$\eqalignno{
\lnabla_\a \Ahat^m_{[1,2]} &= D_\a\Ahat^m_{[1,2]}
-C\dlb\Ahat_\a,\Ahat^m\drb\circ[1,2]&\Firstex\cr
&=D_\a\Ahat^m_{[1,2]} - (k_1\cdot k_2)(\Ahat^1_\a\Ahat^m_2 -
\Ahat^2_\a\Ahat^m_1)\cr
}$$
where we used the first example in \CVTexs.
Therefore the equation of motion of $\Ahat^m_{[1,2]}$ reads
\eqn\EOMex{
D_\a\Ahat^m_{[1,2]} =(\g^m  \What_{[1,2]})_\a + k^m_{12}  \Ahat_{\a\,[1,2]}+
(k_1\cdot k_2)(\Ahat^1_\a\Ahat^m_2 - \Ahat^2_\a\Ahat^m_1)\,.
}

\appendix{C}{Symmetries and deconcatenations of Berends--Giele currents}

\newsubsec\symBG Symmetries of Berends--Giele currents

We have seen on section~\BGmapsec\ that $b(P)$ is a Lie polynomial. A standard result in
the theory of free Lie algebras states that any Lie polynomial is orthogonal
to non-trivial shuffles \Reutenauer. This implies that
\eqn\BGsymEq{
\langle A\shuffle B, b(P)\rangle = 0\,,\quad\forall
A,B\neq\emptyset\quad\len{A}+\len{B}=\len{P}\,,
}
where $\langle\cdot,\cdot\rangle$ is the scalar product of words
and $\shuffle$ is the shuffle product defined in \AdotB\ and \shuffledef, respectively.
A more compact way of stating \BGsymEq\ is through the shorthand
$b(A\shuffle B) =0$.

Using the property \BGsymEq\ it follows that every Berends--Giele current
defined via \BGmap\ is annihilated by proper shuffles, i.e. (note
$\cK_{A\shuffle B} \equiv \sum_{\sigma\in A\shuffle B} \cK_\sigma$)
\eqn\BGexp{
\cK_{A\shuffle B}=0\,,\quad\hbox{$\forall A,B\neq\emptyset$.}
}
Note that the original currents $J^m_{P}$ defined by Berends and Giele in \BGpaper\ were
argued to satisfy $J^m_{A\shuffle B}=0$ in \BGSym. One can show that, in our
conventions, $J^m_P=\cA^m_P$ \BGPS.

\ifig\BGfour{The Berends--Giele current $M_{1234}=\dlb V\drb\circ b(1234)$
according to the map \BGmap.}
{\epsfxsize=0.88\hsize\epsfbox{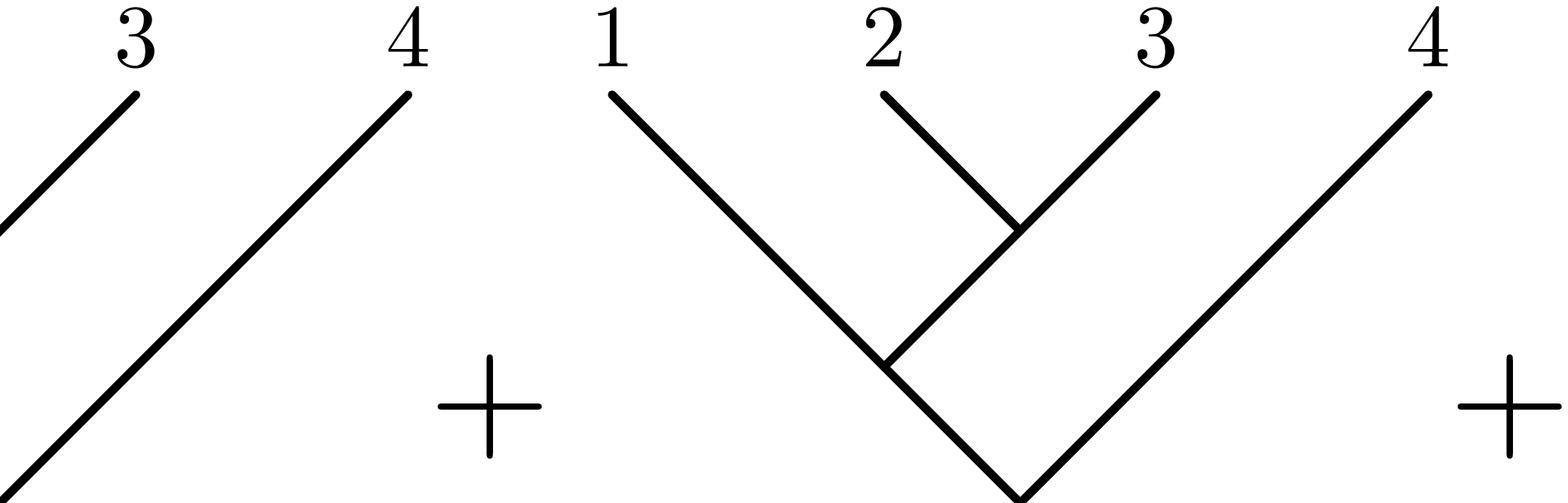}}

\newsubsec\BGeomsec Deconcatenation terms in the equations of motion

The equations of motion of local multiparticle
superfields (see the appendix~\eomapp) contain contact-term corrections with
respect to their single-particle counterparts. When expressed in terms of
Berends--Giele currents, these contact terms corrections are translated to a
deconcatenation structure. For example, the Berends--Giele counterpart of the
local equation of motion
\eqn\locAhatwo{
D_\a\Ahat^m_{[1,2]} =(\g^m  \What_{[1,2]})_\a + k^m_{12}  \Ahat_{\a\,[1,2]}+
(k_1\cdot k_2)(\Ahat^1_\a\Ahat^m_2 - \Ahat^2_\a\Ahat^m_1)\,,
}
is given by
\eqn\nonlocAhatwo{
D_\a\hat {\cal A}^m_{12} =(\g^m  \hat {\cal W}_{12})_\a + k^m_{12}  \hat{\cal
A}^{12}_{\a}+
\sum_{XY=12}\big(\hat{\cal A}^X_\a \hat{\cal A}^m_Y - (X\leftrightarrow Y)\big)\,.
}
These observations can
now be given a universal justification as follows.
If one assigns the superfields
$K$ and $S$ to the contact terms of a Lie polynomial $[P,Q]$ as
\eqn\CVTdef{
C\dlb K,S\drb\circ [P,Q] \equiv \dlb K,S\drb\circ\big(C\circ [P,Q]\big)\,,
}
it follows from \propGoal\  that
\eqn\QBG{
C\dlb K,S\drb\circ b(P) = \sum_{XY=P}\big( \cK_X {\cal S}_Y - (X\leftrightarrow
Y)\big)\,,
}
which demonstrates several deconcatenation formulas of this kind from a local superfield
perspective.
Using the contact-term map $C$ displayed in \Cexamples, the simplest example
applications of \CVTdef\ read
\eqnn\CVTexs
$$\eqalignno{
C\dlb \Ahat_\a,\Ahat^m\drb\circ [1,2] &= (k_1\cdot k_2)(\Ahat_\a^1 \Ahat^m_2 - \Ahat_\a^2 \Ahat^m_1)\,, &\CVTexs\cr
C\dlb V,T\drb\circ [[1,2],3] &= (k_1\cdot k_2)\big(V_{[1,3]} T_2 + V_1 T_{[2,3]} -
V_{[2,3]} T_1 - V_2 T_{[1,3]}\big) \cr
&\quad{}+ (k_{12}\cdot k_3)\big(V_{[1,2]}T_3 - V_3T_{[1,2]}\big)\,.\cr
}$$
In addition, the contact terms generated with the formula \CVTdef\ can be
used to write down the BRST variations of the multiparticle unintegrated $V_P$ for
arbitrary nested Lie bracketings. This generalizes the previous formula valid for
the left-to-right nesting \EOMbbs. More
precisely, the BRST variation can be written as
\eqn\QVPQ{
QV_{[P,Q]} = \half C\dlb V,V\drb\circ[P,Q]\,.
}
For example, using \QVPQ\ one can write down the BRST variation of
$V_{[1,[2,3]]}$ directly,
\eqn\VPQex{
QV_{[1,[2,3]]} = (k_2\cdot k_3)\big(V_{[1,2]}V_3 + V_2V_{[1,3]}\big) + (k_1\cdot
k_{23})V_1V_{[2,3]}\,.
}
Previously one would need to use $V_{[1,[2,3]]}= V_{123} - V_{132}$ before
applying the formula for $QV_P$ for
$P=[[...[p_1,p_2],p_3],...],p_{\len{P}}]\equiv p_1p_2 \ldots p_\len{P}$
given in \oneloopI,
\eqn\QVPo{
QV_{P} = \!\!\sum_{P=XjY\atop \d(Y)=R\otimes S}\!\!(k_{X} \cdot k_j)\,V_{XR}V_{jS}\,.
}
It is worth mentioning that \CofSpecialTopology\ shows the equivalence between
\QVPQ\ and \QVPo.

\appendix{D}{Example applications of the $C$ and $\tilde C$ maps}
\applab\Cexamplesapp

\noindent In this appendix we display some example applications of the $C$ and
$\tilde C$ maps acting over some simple Lie polynomials. These examples help
to elucidate how the algorithms are used, and can be used
to verify that the redefinition formulas arising from the general formulas
match the formulas for the simplest cases that were previously known.

\subsecno=1
\newsubsubsec\Cexsec Examples of the $C$ map

To demonstrate the \recurPQ\ algorithm, the first few expansions generated
from it are
\eqnn\Cexamples
$$\eqalignno{
C\circ 1 &=0 &\Cexamples\cr
C\circ [1,2] &= (k_1\cdot k_2)(1\otimes 2 - 2\otimes 1)\cr
C\circ [[1,2],3] &= (k_1\cdot k_2)\big([1,3]\otimes 2 + 1\otimes [2,3] -
[2,3]\otimes 1 - 2\otimes [1,3]\big) \cr
&\quad{}+ (k_{12}\cdot k_3)\big([1,2]\otimes 3 - 3\otimes [1,2]\big)\cr
C\circ [1,[2,3]] &= (k_2\cdot k_3)\big([1,2]\otimes 3 + 2\otimes [1,3] -
[1,3]\otimes 2 - 3\otimes [1,2]\big) \cr
&\quad{}+ (k_{1}\cdot k_{23})\big(1\otimes [2,3] - [2,3]\otimes 1\big)\cr
C\circ [[[1,2],3],4] &=
(k_1\cdot k_2)\big([[1,3],4]\otimes 2 + [1,3]\otimes [2,4] + [1,4]\otimes [2,3] + 1\otimes [[2,3],4] \cr
&\qquad{}\qquad{}- [[2,3],4]\otimes 1 - [2,3]\otimes [1,4] - [2,4]\otimes [1,3] - 2\otimes [[1,3],4]\big)\cr
&\quad{}+(k_{12}\cdot k_3)\big([[1,2],4]\otimes 3+ [1,2]\otimes [3,4] - [3,4]\otimes [1,2] - 3\otimes [[1,2],4]\big)\cr
&\quad{}+(k_{123}\cdot k_4)\big([[1,2],3]\otimes 4-4\otimes [[1,2],3]\big)\cr
C\circ [[1,[2,3]],4] &=(k_2\cdot k_3)\big([[1,2],4]\otimes 3 + [1,2]\otimes [3,4] + [2,4]\otimes [1,3] + 2\otimes [[1,3],4]\cr
&\qquad{}\qquad{}-[[1,3],4]\otimes 2 - [1,3]\otimes [2,4] - [3,4]\otimes [1,2] - 3\otimes [[1,2],4]\big)\cr
&\quad{}+(k_{1}\cdot k_{23})\big( [1,4]\otimes [2,3] + 1\otimes [[2,3],4] - [[2,3],4]\otimes 1 - [2,3]\otimes [1,4]\big)\cr
&\quad{}+(k_{123}\cdot k_4)\big([1,[2,3]]\otimes 4-4\otimes [1,[2,3]]\big)\cr
C\circ [[1,2],[3,4]] &=(k_1\cdot k_2)\big([1,[3,4]]\otimes 2 + 1\otimes [2,[3,4]] - [2,[3,4]]\otimes 1 - 2\otimes [1,[3,4]]\big) \cr
&\quad{}+(k_3\cdot k_4)\big( [[1,2],3]\otimes 4 + 3\otimes [[1,2],4] - [[1,2],4] \otimes 3 - 4\otimes [[1,2],3]\big) \cr
&\quad{}+(k_{12}\cdot k_{34})\big( [1,2]\otimes [3,4] - [3,4]\otimes [1,2]\big)\cr
C\circ [1,[2,[3,4]]] &=(k_3\cdot k_4)\big( [1,[2,3]]\otimes 4 + [2,3]\otimes [1,4] + [1,3]\otimes [2,4] + 3\otimes [1,[2,4]]\cr
&\qquad{}\qquad{}-[1,[2,4]]\otimes 3 - [2,4]\otimes [1,3] - [1,4]\otimes [2,3] - 4\otimes [1,[2,3]]     \big) \cr
&\quad{}+(k_2\cdot k_{34})\big( [1,2]\otimes [3,4] + 2\otimes [1,[3,4]] - [1,[3,4]]\otimes 2 - [3,4]\otimes [1,2]\big)\cr
&\quad{}+(k_1\cdot k_{234})\big(1\otimes [2,[3,4]] - [2,[3,4]]\otimes 1\big)\cr
C\circ [1,[[2,3],4]] &=(k_2\cdot k_3)\big( [1,[2,4]]\otimes 3 + [1,2]\otimes [3,4] + [2,4]\otimes [1,3] + 2\otimes [1,[3,4]] \cr
&\qquad{}\qquad{}-[1,[3,4]]\otimes 2 - [1,3]\otimes [2,4] - [3,4]\otimes [1,2] - 3\otimes [1,[2,4]]\big)\cr
&\quad{}+(k_{23}\cdot k_4)\big( [1,[2,3]]\otimes 4 + [2,3]\otimes [1,4] - [1,4]\otimes [2,3] - 4\otimes [1,[2,3]]\big)\cr
&\quad{}+(k_1\cdot k_{234})\big(1\otimes [[2,3],4] - [[2,3],4]\otimes 1 \big)\,.
}$$
One application at multiplicity five is given by
\eqnn\justify
$$\eqalignno{
C\circ[[[1,2],3],[4,5]] &= (k_{1}\cdot k_{2})\big(
 1 \otimes  {[ [ 2 , 3 ] , [ 4 , 5 ] ]}
+  {[ 1 , 3 ] \otimes  [ 2 , [ 4 , 5 ] ]} &\justify\cr
&\qquad{}\qquad{} +  {[ 1 , [ 4 , 5 ] ] \otimes  [ 2 , 3 ]}
+  {[ [ 1 , 3 ] , [ 4 , 5 ] ] \otimes  2}
-(1\leftrightarrow2)\Big)\cr
&{}+  (k_{12}\cdot k_{3})\big(
 {[ 1 , 2 ] \otimes  [ 3 , [ 4 , 5 ] ]}
+  {[ [ 1 , 2 ] , [ 4 , 5 ] ] \otimes  3} - ([1,2]\leftrightarrow 3)\big)\cr
&{} +  (k_{123}\cdot k_{45})   \big(
{[ [ 1 , 2 ] , 3 ] \otimes  [ 4 , 5 ]} - ([[1,2],3]\leftrightarrow [4,5])\big)\cr
&{}+  (k_{4}\cdot k_{5})\big(
 4 \otimes  {[ [ [ 1 , 2 ] , 3 ] , 5 ]}
+  {[ [ [ 1 , 2 ] , 3 ] , 4 ] \otimes  5} - (4\leftrightarrow5)\big)\,,
}$$
which, after using the formula \genRedef, reproduces the redefinition (B.2) from \Gauge\ which was
written down without justification.

\newsubsubsec\Ctildesec Examples of the $\tilde C$ map

As an illustration of the $\tilde C$ map, we get
\eqnn\CexamplesA
$$\eqalignno{
\tilde C\circ 1 &=0 &\CexamplesA\cr
\tilde C\circ [1,2] &= 0\cr
\tilde C\circ [[1,2],3] &= (k_1\cdot k_2)\big([1,3]\otimes 2  -
[2,3]\otimes 1 \big) \cr
\tilde C\circ [1,[2,3]] &= (k_2\cdot k_3)\big([1,2]\otimes 3 -
[1,3]\otimes 2 \big) \cr
\tilde C\circ [[[1,2],3],4] &=
(k_1\cdot k_2)\big([[1,3],4]\otimes 2 + [1,4]\otimes [2,3]  -[[2,3],4]\otimes 1 - [2,4]\otimes [1,3] \big)
\cr
&\quad{}+(k_{12}\cdot k_3)\big([[1,2],4]\otimes 3- [3,4]\otimes [1,2] \big)
\cr
\tilde C\circ [[1,[2,3]],4] &=(k_2\cdot k_3)\big([[1,2],4]\otimes 3 + [2,4]\otimes [1,3] -[[1,3],4]]\otimes 2 - [3,4]\otimes [1,2] \big)
\cr
&\quad{}+(k_{1}\cdot k_{23})\big( [1,4]\otimes [2,3] - [[2,3],4]\otimes 1 \big)\cr
\tilde C\circ [[1,2],[3,4]] &=(k_1\cdot k_2)\big([1,[3,4]]\otimes 2  - [2,[3,4]]\otimes 1 \big) \cr
&\quad{}+(k_3\cdot k_4)\big( [[1,2],3]\otimes 4 - [[1,2],4] \otimes 3 \big)\cr
\tilde C\circ [1,[2,[3,4]]] &=(k_3\cdot k_4)\big( [1,[2,3]]\otimes 4 + [1,3]\otimes [2,4] -[1,[2,4]]\otimes 3 -[1,4]\otimes [2,3]     \big) \cr
&\quad{}+(k_2\cdot k_{34})\big( [1,2]\otimes [3,4] - [1,[3,4]]\otimes 2 \big)\cr
\tilde C\circ [1,[[2,3],4]] &=(k_2\cdot k_3)\big( [1,[2,4]]\otimes 3 +[1,2]\otimes [3,4]-[1,[3,4]]\otimes 2 - [1,3]\otimes [2,4] \big)\cr
&\quad{}+(k_{23}\cdot k_4)\big( [1,[2,3]]\otimes 4 - [1,4]\otimes [2,3]\big)\cr
}$$
One application at multiplicity six is given by
\eqnn\CtildeRankSixExample
$$\eqalignno{
\tilde C\circ [[[[1,2],3],[4,5]],6]&=(k_1\cdot
k_2)\big([[[1,3],[4,5]],6]\otimes 2 + [[1,3],6]\otimes [2,[4,5]]&\CtildeRankSixExample\cr
&\qquad{}\qquad{} + [[1,[4,5]],6]\otimes [2,3]+[1,6]\otimes [[2,3],[4,5]] - (1\leftrightarrow 2)\big)\cr
&+(k_{12}\cdot k_3)\big([[[1,2],[4,5]],6]\otimes 3+[[1,2],6]\otimes [3,[4,5]] - ([1,2]\leftrightarrow 3)\big)\cr
&+(k_4\cdot k_5)\big([[[[1,2],3],4],6]\otimes 5 + [4,6]\otimes [[[1,2],3],5] - (4\leftrightarrow 5)\big)\cr
&+(k_{123}\cdot k_{45})\big([[[1,2],3],6]\otimes
[4,5]-([[1,2],3]\leftrightarrow [4,5])\big)\,.
}$$
This will be of particular use in the example discussed in section \Hhatsec.

\appendix{E}{Freedom in defining $H$s}
\applab\HtildeFreedom

\noindent There is considerable freedom in defining the $H$s, arising from the symmetries
within the $H'_{A,B,C}$ terms. These are by construction antisymmetric
in $A$, $B$, and $C$. Furthermore each of the sets of indices will satisfy generalized
Jacobi identities, for instance
\eqnn\Hpjacs
$$\eqalignno{
H'_{123,B,C}+H'_{213,B,C}&=0,&\Hpjacs\cr
H'_{123,B,C}+H'_{231,B,C}+H'_{312,B,C}&=0.
}$$
Also there are a number of other more complex relations between some $H'$ terms,
which can be identified from the condition that $H_{[A,B]}$ satisfies generalized
Jacobi identities in each of $A$ and $B$. For example, we must have that
$\lie_3\circ H_{[123,4]}=0$, $\lie_3\circ H_{[1234,5]}=0$, and
$\lie_4\circ H_{[1234,5]}=0$, and so writing these relations in terms of their
$H'$ expansions, we see that we must have
\eqnn\kernLs
$$\eqalignno{
\lie_3\circ \big(H'_{12,3,4} + H'_{34,1,2}\big) &=0, &\kernLs\cr
\lie_3\circ\big(H'_{123,4,5}-H'_{543,2,1}+H'_{54,3,12}\big)&=0,\cr
\lie_4\circ\big(H'_{123,4,5}-H'_{543,2,1}+H'_{54,3,12}\big)&=0.\cr
}$$
These identities can be described in general with the
formula \genH\ for $H_{[A,B]}$. Consider
$\lie_n\circ H_{[A,B]}$, with $n\leq\len{A}$. One
half of \genH\ will disappear under the action of the $\lie$, as
\eqn\disappearingBitOfH{
\lie_n\circ\left(
\sum_{XjY=\dot a\tilde B}(-1)^{|Y|}H'_{\tilde Y,j,X}
\right)=\lie_n\circ\left(
\sum_{XjY=\tilde B}(-1)^{|Y|}H'_{\tilde Y,j,\dot a X}\right)=0,
}
where in the second sum $X$ is not constrained to be non-empty. The final equality then
just comes from the fact that $H'_{A,B,C}$ is constructed so as to satisfy generalized
Jacobi identities in each of $A$, $B$, and $C$. Using this and \genH\ it then just follows that,
if $\lie_n \circ H_{[A,B]}=0$ for $n\leq\len{A}$, then
\eqn\complexIdentities{
\lie_n\circ\left(
\sum_{XjY=\dot b\tilde A}(-1)^{|Y|}H'_{\tilde Y,j,X}
\right)=0,\;\;\;\;\;\; n<\len{A}
}
for any word $A$ and letterification $\dot b$.

\appendix{F}{BCJ gauge versus Lorenz gauge at multiplicity six}

\noindent The redefinitions for moving from the Lorenz to the BCJ gauge for all
possible topologies at rank six are identified with the usual formula
\genRedef, and are stated below for convenience. We emphasize the typographical
convention of representing a left-to-right nested bracket by its composing letters,
e.g. $\Hhat_{[[[1,2],3],4]}\equiv \Hhat_{1234}$, even though the parent superfields do not obey
BCJ symmetries.
\eqnn\RankSixRedefOne
\eqnn\RankSixRedefTwo
\eqnn\RankSixRedefThree
\eqnn\RankSixRedefFour
\eqnn\RankSixRedefFive
\eqnn\RankSixRedefSix
$$\eqalignno{
A_{[12345,6]}^m &=\hat{A}_{[12345,6]}^m-k_{123456}^m{\hat{H}}_{[12345,6]}\cr
&-(k^1\cdot k^2)\Big(\hat{H}_{13456}\hat{A}_{2}^m+\hat{H}_{1345}\hat{A}_{26}^m+\hat{H}_{1346}\hat{A}_{25}^m+\hat{H}_{1356}\hat{A}_{24}^m\cr
&\;\;\;\;\;\;\;\;\;\;\;\;+\hat{H}_{1456}\hat{A}_{23}^m+\hat{H}_{134}\hat{A}_{256}^m+\hat{H}_{135}\hat{A}_{246}^m+\hat{H}_{136}\hat{A}_{245}^m\cr
&\;\;\;\;\;\;\;\;\;\;\;\;+\hat{H}_{145}\hat{A}_{236}^m+\hat{H}_{146}\hat{A}_{235}^m+\hat{H}_{156}\hat{A}_{234}^m\cr
&\;\;\;\;\;\;\;\;\;\;\;\; -\frac{1}{2}\hat{H}_{134}\hat{H}_{256}k_{256}^m-\frac{1}{2}\hat{H}_{135}\hat{H}_{246}k_{246}^m-\frac{1}{2}\hat{H}_{136}\hat{H}_{245}k_{245}^m\cr
&\;\;\;\;\;\;\;\;\;\;\;\; -\frac{1}{2}\hat{H}_{145}\hat{H}_{236}k_{236}^m-\frac{1}{2}\hat{H}_{146}\hat{H}_{235}k_{235}^m-\frac{1}{2}\hat{H}_{156}\hat{H}_{234}k_{234}^m
\cr
&\;\;\;\;\;\;\;\;\;\;\;\;  - (1\leftrightarrow 2)\Big)\cr
&-(k^{12}\cdot k^3)\Big(\hat{H}_{12456}\hat{A}_{3}^m+\hat{H}_{1245}\hat{A}_{36}^m+\hat{H}_{1246}\hat{A}_{35}^m+\hat{H}_{1256}\hat{A}_{34}^m\cr
&\;\;\;\;\;\;\;\;\;\;\;\; +\hat{H}_{124}\hat{A}_{356}^m+\hat{H}_{125}\hat{A}_{346}^m+\hat{H}_{126}\hat{A}_{345}^m\cr
&\;\;\;\;\;\;\;\;\;\;\;\; -\frac{1}{2}\hat{H}_{124}\hat{H}_{356}k_{356}^m-\frac{1}{2}\hat{H}_{125}\hat{H}_{346}k_{346}^m-\frac{1}{2}\hat{H}_{126}\hat{H}_{345}k_{345}^m\cr
&\;\;\;\;\;\;\;\;\;\;\;\; -\hat{H}_{3456}\hat{A}_{12}^m-\hat{H}_{345}\hat{A}_{126}^m-\hat{H}_{346}\hat{A}_{125}^m-\hat{H}_{356}\hat{A}_{124}^m\cr
&\;\;\;\;\;\;\;\;\;\;\;\; +\frac{1}{2}\hat{H}_{345}\hat{H}_{126}k_{126}^m+\frac{1}{2}\hat{H}_{346}\hat{H}_{125}k_{125}^m+\frac{1}{2}\hat{H}_{356}\hat{H}_{124}k_{124}^m
\Big)\cr
&-(k^{123}\cdot k^4)\Big(\hat{H}_{12356}\hat{A}_{4}^m+\hat{H}_{1235}\hat{A}_{46}^m+\hat{H}_{1236}\hat{A}_{45}^m\cr
&\;\;\;\;\;\;\;\;\;\;\;\;+\hat{H}_{123}\hat{A}_{456}^m-\hat{H}_{456}\hat{A}_{123}^m\cr
&\;\;\;\;\;\;\;\;\;\;\;\;
-\frac{1}{2}\hat{H}_{123}\hat{H}_{456}k_{456}^m+\frac{1}{2}\hat{H}_{456}\hat{H}_{123}k_{123}^m\Big)\cr
&-(k^{1234}\cdot k^5)\Big(\hat{H}_{12346}\hat{A}_{5}^m+\hat{H}_{1234}\hat{A}_{56}^m\Big)\cr
&-(k^{12345}\cdot k^6)\hat{H}_{12345}\hat{A}_6^m & \RankSixRedefOne
\cr
A^m_{[1234,56]} &=\hat{A}_{[1234,56]}^m-k_{123456}^m{\hat{H}}_{[1234,56]}\cr
 &-(k^1\cdot k^2)\Big( \hat{H}_{[1,56]}\hat{A}_{234}^m + \hat{H}_{[13,56]}\hat{A}_{24}^m+\hat{H}_{[14,56]}\hat{A}_{23}^m \cr
&\;\;\;\;\;\;\;\;\;\;\;\;  + \hat{H}_{[134,56]}\hat{A}_{2}^m+ \hat{H}_{134}\hat{A}_{[2,56]}^m\cr
&\;\;\;\;\;\;\;\;\;\;\;\;  - \frac{1}{2}\hat{H}_{134}\hat{H}_{[2,56]}k_{256}^m-\frac{1}{2}\hat{H}_{[1,56]}\hat{H}_{234}k_{234}^m
-(1\leftrightarrow 2)\Big) \cr
&-(k^{12}\cdot k^3)\Big(\hat{H}_{[12,56]}\hat{A}_{34}^m+\hat{H}_{[124,56]}\hat{A}_{3}^m+\hat{H}_{124}\hat{A}_{[3,56]}^m\cr
&\;\;\;\;\;\;\;\;\;\;\;\; - \hat{H}_{[3,56]}\hat{A}_{124}^m-\hat{H}_{[34,56]}\hat{A}_{12}^m\cr
&\;\;\;\;\;\;\;\;\;\;\;\; -\frac{1}{2}\hat{H}_{124}\hat{H}_{[3,56]}k_{356}^m+ \frac{1}{2}\hat{H}_{[3,56]}\hat{H}_{124}k_{124}^m
\Big)\cr
&- (k^{123}\cdot k^4)\Big(\hat{H}_{[123,56]}\hat{A}_4^m+\hat{H}_{123}\hat{A}^m_{[4,56]}-   \hat{H}_{[4,56]}\hat{A}_{123}^m  \cr
&\;\;\;\;\;\;\;\;\;\;\;\; -\frac{1}{2}\hat{H}_{123}\hat{H}_{[4,56]}k_{456}^m+  \frac{1}{2} \hat{H}_{[4,56]}\hat{H}_{123}k_{123}^m  ) \cr
&- (k^{1234}\cdot k^{56})(\hat{H}_{1234}\hat{A}_{56}^m\Big) \cr
&+(k^5\cdot k^6)\Big(\hat{H}_{[5,1234]}\hat{A}_6^m-(5\leftrightarrow 6 )\Big) & \RankSixRedefTwo
\cr
A^m_{[123,456]} &=\hat{A}_{[123,45,6]}^m-k_{123456}^m{\hat{H}}_{[123,456]}\cr
&-(k^1\cdot
k^2)\Big(\hat{H}_{[1,456]}\hat{A}_{23}^m+\hat{H}_{[13,456]}\hat{A}_{2}^m-(1\leftrightarrow 2)\Big)\cr
&-(k^{12}\cdot
k^{3})\Big(\hat{H}_{[12,456]}\hat{A}_3^m-\hat{H}_{[3,456]}\hat{A}_{12}^m\Big)\cr
&-(k^{123}\cdot k^{456})\Big(\hat{H}_{123}\hat{A}_{456}^m-\hat{H}_{456}\hat{A}_{123}^m \cr
&\;\;\;\;\;\;\;\;\;\;\;\;
-\frac{1}{2}\hat{H}_{123}\hat{H}_{456}k_{456}^m+\frac{1}{2}\hat{H}_{456}\hat{H}_{123}k_{123}^m
\Big)\cr
&+(k^4\cdot
k^5)\Big(\hat{H}_{[4,123]}\hat{A}_{56}^m+\hat{H}_{[46,123]}\hat{A}_{5}^m-(4\leftrightarrow 5)\Big)\cr
&+(k^{45}\cdot k^{6})\Big(\hat{H}_{[45,123]}\hat{A}_6^m-\hat{H}_{[6,123]}\hat{A}_{45}^m\Big)& \RankSixRedefThree
\cr
   A^m_{[[12,34],56]} &=\hat{A}_{[[12,34],56]}^m-k_{123456}^m{\hat{H}}_{[[12,34],56]}\cr
&- (k^1\cdot k^2)\Big(\hat{H}_{[1,34]}\hat{A}_{[2,56]}^m+\hat{H}_{[1,56]}\hat{A}_{[2,34]}^m+\hat{H}_{[[1,34],56]}\hat{A}_{2}^m \cr
&\;\;\;\;\;\;\;\;\;\;\;\;
-\frac{1}{2}\hat{H}_{[1,34]}\hat{H}_{[2,56]}k_{256}^m-\frac{1}{2}\hat{H}_{[1,56]}\hat{H}_{[2,34]}k_{234}^m
-(1\leftrightarrow 2)\Big)\cr
&+ (k^3\cdot k^4)\Big(\hat{H}_{[3,12]}\hat{A}_{[4,56]}^m+\hat{H}_{[3,56]}\hat{A}_{[4,12]}^m+\hat{H}_{[[3,12],56]}\hat{A}_{4}^m \cr
&\;\;\;\;\;\;\;\;\;\;\;\;
-\frac{1}{2}\hat{H}_{[3,12]}\hat{H}_{[4,56]}k_{456}^m-\frac{1}{2}\hat{H}_{[3,56]}\hat{H}_{[4,12]}k_{124}^m
-(3\leftrightarrow 4)\Big)\cr
&- (k^{12}\cdot
k^{34})\Big(\hat{H}_{[12,56]}\hat{A}_{34}^m-\hat{H}_{[34,56]}\hat{A}_{12}^m\Big)\cr
&- (k^{1234}\cdot k^{56})\Big(\hat{H}_{[12,34]}\hat{A}_{56}^m\Big)\cr
&+ (k^5\cdot
k^6)\Big(\hat{H}_{[[12,34],6]}\hat{A}_5^m-\hat{H}_{[[12,34],5]}\hat{A}_6^m\Big)& \RankSixRedefFour
\cr
   A_{[[123,45],6]}^m &=\hat{A}_{[[123,45],6]}^m-k_{123456}^m{\hat{H}}_{[[123,45],6]}\cr
&-(k^1\cdot k^2)\Big(\hat{H}_{[1,45]}\hat{A}^m_{236}+\hat{H}_{136}\hat{A}_{[2,45]}^m+\hat{H}_{[[1,45],6]}\hat{A}_{23}^m\cr
&\;\;\;\;\;\;\;\;\;\;\;\; +\hat{H}_{[13,45]}\hat{A}^m_{26}+\hat{H}_{[[13,45],6]}\hat{A}_2^m\cr
&\;\;\;\;\;\;\;\;\;\;\;\;-\frac{1}{2}\hat{H}_{[1,45]}\hat{H}_{236}k_{236}^m-\frac{1}{2}\hat{H}_{136}\hat{H}_{[2,45]}k_{245}^m-(1\leftrightarrow
2)\Big)\cr
&-(k^{12}\cdot k^3)\Big(\hat{H}_{126}\hat{A}^m_{[3,45]}+\hat{H}_{[12,45]}\hat{A}_{36}^m+\hat{H}_{[[12,45],6]}\hat{A}_3^m\cr
&\;\;\;\;\;\;\;\;\;\;\;\; - \hat{H}_{[3,45]}\hat{A}^m_{126}-\hat{H}_{[[3,45],6]}\hat{A}_{12}^m \cr
&\;\;\;\;\;\;\;\;\;\;\;\;
-\frac{1}{2}\hat{H}_{126}\hat{H}_{[3,45]}k_{345}^m+\frac{1}{2}
\hat{H}_{[3,45]}\hat{H}_{126}k_{126}^m\Big)\cr
&+ (k^4\cdot
k^5)\Big(\hat{H}_{[4,123]}\hat{A}_{56}^m+\hat{H}_{[[4,123],6]}\hat{A}_5^m-(4\leftrightarrow 5)\Big)\cr
&-(k^{123}\cdot k^{45})\Big(\hat{H}_{1236}\hat{A}_{45}^m+\hat{H}_{123}\hat{A}_{456}^m-\hat{H}_{456}\hat{A}_{123}^m\cr
&\;\;\;\;\;\;\;\;\;\;\;\;
-\frac{1}{2}\hat{H}_{123}\hat{H}_{456}k_{456}^m+\frac{1}{2}\hat{H}_{456}\hat{H}_{123}k_{123}^m\Big)\cr
&-(k^{12345}\cdot k^6)\Big(\hat{H}_{[123,45]}\hat{A}_6^m\Big)& \RankSixRedefFive
\cr
A_{[[[12,34],5],6]}^m &=\hat{A}_{[[[12,34],5],6]}^m-k_{123456}^m{\hat{H}}_{[[[12,34],5],6]}\cr
&-(k^1\cdot k^2)\Big(\hat{H}_{156}\hat{A}_{[2,34]}^m+\hat{H}_{[1,34]}\hat{A}_{256}^m+\hat{H}_{[[1,34],6]}\hat{A}_{25}^m\cr
&\;\;\;\;\;\;\;\;\;\;\;\; +\hat{H}_{[[1,34],5]}\hat{A}_{26}^m+\hat{H}_{[[[1,34],5],6]}\hat{A}_{2}^m \cr
&\;\;\;\;\;\;\;\;\;\;\;\;
-\frac{1}{2}\hat{H}_{156}\hat{H}_{[2,34]}k_{234}^m-\frac{1}{2}\hat{H}_{[1,34]}\hat{H}_{256}k_{256}^m-(1\leftrightarrow 2)\Big)\cr
&+(k^3\cdot k^4)\Big(\hat{H}_{356}\hat{A}_{[4,12]}^m+\hat{H}_{[3,12]}\hat{A}_{456}^m+\hat{H}_{[[3,12],6]}\hat{A}_{45}^m\cr
&\;\;\;\;\;\;\;\;\;\;\;\; +\hat{H}_{[[3,12],5]}\hat{A}_{46}^m+\hat{H}_{[[[3,12],5],6]}\hat{A}_{4}^m \cr
&\;\;\;\;\;\;\;\;\;\;\;\;-\frac{1}{2}\hat{H}_{356}\hat{H}_{[4,12]}k_{124}^m-\frac{1}{2}\hat{H}_{[3,12]}\hat{H}_{456}k_{456}^m-(3\leftrightarrow
4)\Big)\cr
&-(k^{12}\cdot k^{34})\Big(\hat{H}_{1256}\hat{A}_{34}^m+\hat{H}_{126}\hat{A}_{345}^m+\hat{H}_{125}\hat{A}_{346}^m\cr
&\;\;\;\;\;\;\;\;\;\;\;\;
-\frac{1}{2}\hat{H}_{126}\hat{H}_{345}k_{345}^m-\frac{1}{2}\hat{H}_{125}\hat{H}_{346}k_{346}^m
-(12\leftrightarrow 34)\Big)\cr
&-(k^{1234}\cdot
k^5)\Big(\hat{H}_{[12,34]}\hat{A}_{56}^m+\hat{H}_{[[12,34],6]}\hat{A}_{5}^m\Big)\cr
&-(k^{12345}\cdot k^6)\Big(\hat{H}_{[[12,34],5]}\hat{A}_6^m\Big)& \RankSixRedefSix
\cr
}$$
where the redefinition terms $\hat H_P$ are defined so as to enforce generalized
Jacobi identities upon superfields, and can be identified most easily with
repeated use of \HhatDef\ and \genH\ - \HABCdef.

The rank six Berends-Giele currents are found using the Berends-Giele map
\BGmap, and for a general superfield $K$ is
\eqnn\RankSixBGCurrent
$$\eqalignno{
       s_{123456}{\cal K}_{123456}&= \frac{K_{[[[[[1,2],3],4],5],6]}}{s_{12}s_{123}s_{1234}s_{12345}}
       + \frac{K_{[[[[1,[2,3]],4],5],6]}}{s_{123}s_{1234}s_{12345}s_{23}}
       + \frac{K_{[[[[1,2],[3,4]],5],6]}}{s_{12}s_{1234}s_{12345}s_{34}}
       + \frac{K_{[[[[1,2],3],[4,5]],6]}}{s_{12}s_{123}s_{12345}s_{45}}\cr
       &+ \frac{K_{[[[[1,2],3],4],[5,6]]}}{s_{12}s_{123}s_{1234}s_{56}}
       + \frac{K_{[[[1,[[2,3],4]],5],6]}}{s_{1234}s_{12345}s_{23}s_{234}}
       + \frac{K_{[[[1,[2,[3,4]]],5],6]}}{s_{1234}s_{12345}s_{234}s_{34}}
       + \frac{K_{[[[1,[2,3]],[4,5]],6]}}{s_{123}s_{12345}s_{23}s_{45}}\cr
       &+ \frac{K_{[[[1,[2,3]],4],[5,6]]}}{s_{123}s_{1234}s_{23}s_{56}}
       + \frac{K_{[[[1,2],[[3,4],5]],6]}}{s_{12}s_{12345}s_{34}s_{345}}
       + \frac{K_{[[[1,2],[3,[4,5]]],6]}}{s_{12}s_{12345}s_{345}s_{45}}
       + \frac{K_{[[[1,2],[3,4]],[5,6]]}}{s_{12}s_{1234}s_{34}s_{56}}\cr
       &+ \frac{K_{[[[1,2],3],[[4,5],6]]}}{s_{12}s_{123}s_{45}s_{456}}
       + \frac{K_{[[[1,2],3],[4,[5,6]]]}}{s_{12}s_{123}s_{456}s_{56}}
       + \frac{K_{[[1,[[[2,3],4],5]],6]}}{s_{12345}s_{23}s_{234}s_{2345}}
       + \frac{K_{[[1,[[2,[3,4]],5]],6]}}{s_{12345}s_{234}s_{2345}s_{34}}\cr
       &+ \frac{K_{[[1,[[2,3],[4,5]]],6]}}{s_{12345}s_{23}s_{2345}s_{45}}
       + \frac{K_{[[1,[[2,3],4]],[5,6]]}}{s_{1234}s_{23}s_{234}s_{56}}
       + \frac{K_{[[1,[2,[[3,4],5]]],6]}}{s_{12345}s_{2345}s_{34}s_{345}}
       + \frac{K_{[[1,[2,[3,[4,5]]]],6]}}{s_{12345}s_{2345}s_{345}s_{45}}\cr
       &+ \frac{K_{[[1,[2,[3,4]]],[5,6]]}}{s_{1234}s_{234}s_{34}s_{56}}
       + \frac{K_{[[1,[2,3]],[[4,5],6]]}}{s_{123}s_{23}s_{45}s_{456}}
       + \frac{K_{[[1,[2,3]],[4,[5,6]]]}}{s_{123}s_{23}s_{456}s_{56}}
       + \frac{K_{[[1,2],[[[3,4],5],6]]}}{s_{12}s_{34}s_{345}s_{3456}}\cr
       &+ \frac{K_{[[1,2],[[3,[4,5]],6]]}}{s_{12}s_{345}s_{3456}s_{45}}
       + \frac{K_{[[1,2],[[3,4],[5,6]]]}}{s_{12}s_{34}s_{3456}s_{56}}
       + \frac{K_{[[1,2],[3,[[4,5],6]]]}}{s_{12}s_{3456}s_{45}s_{456}}
       + \frac{K_{[[1,2],[3,[4,[5,6]]]]}}{s_{12}s_{3456}s_{456}s_{56}}\cr
       &+ \frac{K_{[1,[[[[2,3],4],5],6]]}}{s_{23}s_{234}s_{2345}s_{23456}}
       + \frac{K_{[1,[[[2,[3,4]],5],6]]}}{s_{234}s_{2345}s_{23456}s_{34}}
       + \frac{K_{[1,[[[2,3],[4,5]],6]]}}{s_{23}s_{2345}s_{23456}s_{45}}
       + \frac{K_{[1,[[[2,3],4],[5,6]]]}}{s_{23}s_{234}s_{23456}s_{56}}\cr
       &+ \frac{K_{[1,[[2,[[3,4],5]],6]]}}{s_{2345}s_{23456}s_{34}s_{345}}
       + \frac{K_{[1,[[2,[3,[4,5]]],6]]}}{s_{2345}s_{23456}s_{345}s_{45}}
       + \frac{K_{[1,[[2,[3,4]],[5,6]]]}}{s_{234}s_{23456}s_{34}s_{56}}
       + \frac{K_{[1,[[2,3],[[4,5],6]]]}}{s_{23}s_{23456}s_{45}s_{456}}\cr
       &+ \frac{K_{[1,[[2,3],[4,[5,6]]]]}}{s_{23}s_{23456}s_{456}s_{56}}
       + \frac{K_{[1,[2,[[[3,4],5],6]]]}}{s_{23456}s_{34}s_{345}s_{3456}}
       + \frac{K_{[1,[2,[[3,[4,5]],6]]]}}{s_{23456}s_{345}s_{3456}s_{45}}
       + \frac{K_{[1,[2,[[3,4],[5,6]]]]}}{s_{23456}s_{34}s_{3456}s_{56}}\cr
       &+ \frac{K_{[1,[2,[3,[[4,5],6]]]]}}{s_{23456}s_{3456}s_{45}s_{456}}
       + \frac{K_{[1,[2,[3,[4,[5,6]]]]]}}{s_{23456}s_{3456}s_{456}s_{56}}. &\RankSixBGCurrent
}$$
Verifying that the redefinitions  \RankSixRedefOne\ - \RankSixRedefSix\ amount to a gauge transformation
in the Berends-Giele currents means plugging them into the above, and checking that in the resulting
expression the Mandelstams cancel perfectly and the formula \bcjsixgauge, which
has the form of a gauge transformation, is produced.

Clearly this calculation requires considerable effort, but it
has been performed and the result works as it should. A more efficient alternative
approach based on \propGoal\ of Proposition~1 is possible though, and works
as follows. We begin with the
definition of the BG current, ${\cal A}^{m,BCJ}_{123456}=\dlb A^m\drb\circ b(123456)$.
Using the general form of the gauge transformation \genRedef\
we see that this is just
\eqn\ABCJsix{
{\cal A}^{m,BCJ}_{123456}=\dlb \hat A^m\drb\circ b(123456) - C\dlb \hat H,L_2\circ \hat A^m\drb\circ b(123456) - \dlb \hat H^m\drb\circ b(123456),
}
which by \BGmap\ and \propGoal\ is just
$$\eqalignno{
{\cal A}^{m,BCJ}_{123456}&={\cal A}^{m,L}_{123456} - k^{123456}_m{\cal H}_{123456}
- \dlb \hat H,L_2\circ \hat A^m\drb\circ\!\!\!\!\!\!\! \sum_{XY=12...6}\Big(b(X)\otimes b(Y)-b(Y)\otimes b(X)\Big)\cr
&={\cal A}^{m,L}_{123456} - k^{123456}_m{\cal H}_{123456}
-\!\!\!\!\!\!\! \sum_{XY=12...6}\!\!\!\!\Big({\cal H}_X \dlb L_2\circ \hat A^m\drb \circ b(Y)-{\cal H}_Y \dlb L_2\circ \hat A^m\drb\circ b(X)\Big).
}$$
Completing another round of the same sort of calculation on the $\dlb L_2\circ \hat A^m\drb$ terms
yields\foot{Note there are no $L_3$ terms in the below. These have been omitted intentionally as any
such terms would be of the form $\sum_{XYZ=12...6}H_XH_YA_Z$, and since each $H$ requires at least three indices to be non-zero all terms of this form will be zero.}
\eqnn\nonLinsix
$$\eqalignno{
{\cal A}^{m,BCJ}_{123456}
={\cal A}^{m,L}_{123456} &- k^{123456}_m{\cal H}_{123456}-
\sum_{XY=12...6}\Big({\cal H}_X {\cal A}^{m,L}_Y-{\cal H}_Y {\cal A}^{m,L}_X\Big) &\nonLinsix\cr
&+\frac{1}{2}\sum_{XY=12...6}\Big({\cal H}_X {\cal H}_Yk_Y^m-{\cal H}_Y{\cal H}_Xk_X^m\Big)\,,.
}$$
This is then just \bcjsixgauge, as was desired. By a similar argument it could
be shown that all redefinitions produced by \genRedef\ have the form of a gauge
transformation.

\listrefs
\bye